\documentclass[manuscript,screen,nonacm]{acmart}%
\AtBeginDocument{%
  }

\setcopyright{none}

\copyrightyear{2027}
\acmYear{2027}
\acmDOI{XXXXXXX.XXXXXXX}
\acmConference[CHI '27]{2027 CHI Conference on Human Factors in Computing Systems
}{May 10--14, 2027}{Pittsburgh, PA}
\acmISBN{978-1-4503-XXXX-X/2018/06}

\usepackage{hyperref}
\usepackage{enumitem}
\usepackage{graphicx}
\usepackage{subcaption}
\usepackage{booktabs}
\usepackage{multirow}
\usepackage{longtable}

\usepackage[table]{xcolor}
\usepackage{tabularx}
\usepackage{array}
\usepackage{arydshln}
\usepackage{listings}
\usepackage{xcolor}

\lstdefinestyle{promptstyle}{
    basicstyle=\ttfamily\small,
    numbers=left,                  %
    numberstyle=\tiny\color{gray}, %
    stepnumber=1,                  %
    firstnumber=1,                 %
    numbersep=10pt,                %
    frame=lines,                    %
    breaklines=true,               %
    breakatwhitespace=true,
    showstringspaces=false,
    tabsize=2,
    xleftmargin=1.5em              %
}

\definecolor{needMet}{HTML}{009901}
\definecolor{needNotMet}{HTML}{F56B00}
\newcommand{\metneed}[1]{\textcolor{needMet}{\checkmark~#1}}
\newcommand{\unmetneed}[1]{\textcolor{needNotMet}{?~#1}}

\newcolumntype{L}[1]{>{\raggedright\arraybackslash}p{#1}}
\newcolumntype{Y}{>{\raggedright\arraybackslash}X}

\begin{document}

\title[Explanation Navigator]{Explanation Navigator: Rectifying Out-of-Scope Human Interpretations of Leaky AI Explanations through Conversational Guidance}%
\author{Yueqing Xuan}
\email{yueqing.xuan@rmit.edu.au}
\orcid{0000-0002-9365-8949}
\affiliation{%
  \institution{ARC Centre of Excellence for Automated Decision-Making and Society, School of Computing Technologies, RMIT University}
  \country{Australia}
}

\author{Kacper Sokol}
\affiliation{
  \institution{Universit\`{a} della Svizzera italiana}
  \city{Lugano}
  \country{Switzerland}
}
\email{kacper.sokol@usi.ch}
\orcid{0000-0002-9869-5896}

\author{Danula Hettiachchi}
\email{danula.hettiachchi@rmit.edu.au}
\orcid{0000-0003-3875-5727}
\affiliation{%
  \institution{ARC Centre of Excellence for Automated Decision-Making and Society, School of Computing Technologies, RMIT University}
  \country{Australia}
}

\renewcommand{\shortauthors}{Xuan et al.}

\begin{abstract}

As explanations of artificial intelligence systems proliferate, their recipients must grasp not only what they convey but also recognise what they cannot. We conducted an interview study with nine participants to examine how explainees reason when their information needs exceed the scope of available explanations. Participants often unwittingly confabulated explanatory insights when relevant information was missing from the explanations, not recognising the inherent limitations thereof. We characterise such explanations as \emph{leaky explanations} -- simplifications that strive to hide complexity yet whose correct interpretation hinges on understanding of the concealed details. To address out-of-scope interpretations we propose \emph{Explanation Navigator}: a conversational interaction framework that detects mismatches between users' information needs and explanations' content, elucidating pertinent yet implicit details and providing complementary explanations for unmet information needs. An online study with 316 participants showed that our approach allowed explainees to recognise and rectify confabulated explanatory insights, guiding them towards developing correct understanding.%

\end{abstract}



\maketitle

\section{Introduction}

Artificial intelligence (AI) is increasingly deployed in high-stakes domains, motivating the development of explainable AI (XAI) techniques that help people understand how such tools operate~\cite{guidotti2018survey}. XAI methods differ in the aspects of model behaviour they expose and the purposes they support~\cite{binns2018s}. For example, local feature importance explanations describe how features contribute to a particular prediction, whereas counterfactual explanations identify feature changes associated with an alternative outcome~\cite{dwivedi2023explainable}. %

Human-centred XAI research has focused on whether AI explanations improve users' understanding of, trust towards or fairness perception of AI systems~\cite{hoffman2023measures,van2021effect,schoeffer2022there,ma2025towards}. A related line of work also investigates whether users correctly understand the explanations themselves~\cite{kaur2020interpreting,collaris2022characterizing}. As \citet{jacovi2023diagnosing} argue, the usefulness of an explanation depends not only on the information it contains but also on what the explainee derives from it. A correct interpretation of an explanation therefore requires users not only to understand the information it provides but also to recognise what information remains unspecified~\cite{xuan2025comprehension}.

Consider a user viewing a local feature importance explanation of an AI prediction. They may correctly identify which features contributed the most to the prediction, yet incorrectly infer that a negatively contributing feature should be increased to improve the outcome. The issue is therefore not a failure to understand the information shown but a tendency to confabulate information that the explanation does not actually communicate. %

Prior research has documented many instances of such divergence between the intended semantics of explanations and users' interpretations. For example, users may generalise instance-specific explanations, i.e., local explanations, to unsupported beliefs about aggregate model behaviour~\cite{chromik2021think,bove2022contextualization}. %
Existing work explains such misinterpretations to user-side factors like cognitive biases~\cite{kliegr2021review}, the illusion of explanatory depth~\cite{chromik2021think,rozenblit2002misunderstood} as well as bounded rationality~\cite{kaur2024interpretability}. Other conceptual work has identified potential explanation-side sources of failure, including representational or causal components that are omitted from the explanatory narrative~\cite{jacovi2023diagnosing,bove2024explanations,barocas2020hidden}.

However, less is known about how explanation-side representations and user-side reasoning interact to produce misunderstanding in practice. 
In particular, when users seek information that an explanation does not provide, do they recognise that this information is unavailable, or do they reason beyond the explanation by introducing unfounded assumptions about how it works? Our first research question (\textbf{RQ1}) therefore asks: \emph{How does misunderstanding of an explanation arise when users seek information that the explanation does not provide?}
Elucidating this distinction has implications for how such misunderstandings should be rectified. If the problem is primarily that the current explanation does not contain the sought information, then providing additional explanations may be sufficient. If, however, users have already formed an unsupported interpretation of what the original explanation can establish, additional information alone may be insufficient to revise that interpretation.

This distinction also becomes increasingly important as XAI systems move beyond single explanations and towards dashboards, interactive interfaces or conversational systems that provide access to multiple forms of explanatory information~\cite{bertrand2023selective,slack2023explaining}. Conversational XAI systems, in particular, can identify users' evolving information needs and provide corresponding explanations~\cite{mindlin2024measuring,he2025conversational}. However, these systems are commonly evaluated in terms of whether their explanations improve users' understanding of the underlying AI model. It remains unclear if access to additional explanations can correct a misunderstanding that has formed before based on a particular explanation. Our \textbf{RQ2} therefore asks: \emph{Can additional explanations, provided through dashboards or conversational interaction, repair an existing misunderstanding of an explanation?}

We investigated RQ1 and RQ2 through a semi-structured interview study with nine participants. 
During the study, participants examined three local AI explanations individually and were asked to extract different pieces of information from them, including both insights that the explanations could support and those that they could not. %
They were then given access to additional explanations through a dashboard and a conversational XAI system to examine whether these misunderstandings could be rectified.  
We found that participants generally interpreted information correctly when it was explicitly supported by an explanation. However, when they sought information that was unavailable, they often confabulated plausible assumptions in place of knowledge about the explanatory scope to construct an answer. For example, participants treated characteristics of individual conditional expectation (ICE)~\citep{goldstein2015peeking} curves as evidence of feature importance. %

We attribute this phenomenon to such explanations being \textbf{leaky explanations}. AI explanations abstract away details of their construction -- as well as the operation of the underlying predictive models -- yet these hidden properties constrain what can be correctly inferred from them, i.e., their explanatory scope. When users seek information beyond this scope, they fill in these hidden details with their own assumptions about how the explanations work and use those assumptions to infer an answer. When these assumptions do not align with the explanations' underlying mechanisms, the resulting interpretations may appear plausible while at the same time extending beyond what the explanation can legitimately support.

Our study also found that providing additional explanations through either the dashboard or conventional conversational explainer did not reliably rectify these interpretations. 
This motivates our third research question (\textbf{RQ3}): \emph{How should an interactive explainer support users in recognising and rectifying from unsupported interpretations?} 
To address RQ3, we propose the \textbf{Explanation Navigator}: a conversational interaction framework that mediates the link between the users' information needs and the scope of the explanation currently being interpreted. When the requested information exceeds what the explanation can substantiate, %
the system makes the relevant scope explicit, explains why the inference is unsupported and then provides a complementary explanation capable of addressing the outstanding information needs. In this way, our Explanation Navigator aims to rectify the users' interpretation of the current explanation while still providing the information they seek.

We evaluated the Explanation Navigator in an online study with 316 participants using local feature importance and ICE explanations. %
We measured both participants' comprehension of information provided by the explanation as well as their ability to recognise unsupported inferences. The Explanation Navigator significantly improved out-of-scope information recognition for both explanation types, demonstrating the benefit of combining clarification of the current explanation with access to complementary explanatory information.

In summary, our work makes three contributions:
\begin{itemize}
\item We empirically examine how users reason when their information needs exceed what an explanation offers. We identify a reasoning process in which users confabulate assumptions about how an explanation is generated and insights about what information it describes. %
We attribute this process %
to AI explanations being \emph{leaky}. 

\item We show that access to additional explanations through an XAI dashboard and conversational system does not necessarily rectify an existing misunderstanding. We identify two corrective components: making the scope of the relevant explanation explicit and providing a complementary explanation.

\item We propose the \emph{Explanation Navigator}, which implements these corrective mechanisms to support users in rectifying out-of-scope interpretations; we then demonstrate its effectiveness with a quantitative study.
\end{itemize}

\section{Related Work}

Our work builds on prior research on diverse human-centred XAI methods and the heterogeneous information needs of their users, a review of which follows.

\subsection{AI Explanations and User Information Needs}

Explainable AI encompasses a wide range of methods intended to make AI predictions and model behaviour more intelligible to people~\cite{retzlaff2024post,byrne2023good}. Different explanation methods expose different information about the AI model and therefore have different explanatory scopes~\cite{lipton2018mythos,doshi2017towards,mohseni2021multidisciplinary}. For example, local feature importance explanations describe feature relevance for a particular prediction, counterfactual explanations identify alternative feature values associated with a different outcome and global explanations characterise general model behaviour~\cite{christoph2020interpretable}. Consequently, any single explanation provides only a selective view of an AI model rather than a complete account of its operation~\cite{miller2019explanation}.

The explanatory scope of a method is shaped by the technical properties of the process responsible for generating the explanation. These properties determine what aspects of model behaviour the explanation conveys and, consequently, what conclusions can be validly drawn from it. For example, counterfactual explanations depend on choices concerning actionability, feature relationships and the underlying AI model~\cite{barocas2020hidden,verma2024counterfactual}, whereas feature attribution methods differ in locality, additivity or faithfulness~\cite{ribeiro2016should,lundberg2017unified,ribeiro2018anchors}. However, many of these properties are not apparent from the explanation itself, even though they constrain how the explanation should be interpreted.

At the same time, users approach explanations with diverse goals. Prior work showed that explanatory needs vary across users, expertise levels, tasks and contexts~\cite{lim2009and,cai2019hello,liao2020questioning,wang2021explanations,liao2022connecting,kim2024xai,rong2023towards}. For example, clinicians may seek both case-specific reasons as well as broader information about model behaviour~\cite{cai2019hello}, while other users may seek explanations to understand a prediction, calibrate trust, identify possible actions or support task learning~\cite{kim2024xai}. Thus, there is unlikely to be a universally preferable explanation; different information needs require different explanation methods~\cite{sokol2020one}.

This has motivated XAI systems that provide multiple explanation types or adapt explanations as users' information needs evolve~\cite{bhattacharya2024exmos,he2025conversational,slack2023explaining}. These systems primarily focus on satisfying the user's next information need by connecting it to an appropriate explanation. However, users may already have formed an unsupported interpretation of the explanation currently in front of them without recognising that the inference they made exceeds its scope~\cite{xuan2025comprehension}. In such cases, providing another explanation may answer the next question without correcting the original misunderstanding. Our work therefore focuses on the gap between \emph{what an explanation supports} and \emph{what users believe they can infer from it}.

\subsection{User Misunderstanding of AI Explanations}

Human-centred evaluation of AI explanations often examines whether explanations help users understand the underlying AI model, for example by testing whether users can simulate model behaviour~\cite{rong2023towards,bobek2025user,cheng2019explaining}. Other work evaluates whether explanations help users detect model errors~\cite{wang2021explanations}, improve decision-making accuracy~\cite{bansal2021does,alufaisan2021does} or support calibrated trust and appropriate reliance on AI recommendations~\cite{zhang2020effect,buccinca2021trust,naiseh2023different}.

However, these evaluation strategies do not directly establish whether users correctly understand the explanation itself. As \citet{jacovi2023diagnosing} argued, explanation usefulness depends not only on the information presented but also on what users derive from it. A growing body of work therefore examined users' understanding of explanations themselves~\cite{rong2023towards} and consistently showed that users' interpretations can diverge from the intended meaning of the explanation. Among data scientists, \citet{kaur2020interpreting} found widespread misuse of interpretability tools, while \citet{collaris2022characterizing} found that practitioners' expectations of local feature importance did not consistently align with the assumptions of the underlying methods. Similar problems occurred among non-experts: users may generalise local explanations into global beliefs~\cite{chromik2021think,bove2022contextualization}, infer more from saliency explanations than they actually support~\cite{alqaraawi2020evaluating} or assign unintended meaning to feature attribution visualisations~\cite{zhang2025may}. \citet{xuan2025comprehension} further showed that highly comprehensible explanations can still produce inaccurate interpretations.

Some work explained these misunderstandings through user-side reasoning. \citet{kliegr2021review} showed how cognitive biases can shape the interpretation of rule-based models, while \citet{chromik2021think} drew on the illusion of explanatory depth~\cite{rozenblit2002misunderstood} to argue that simplified local explanations can lead users to overestimate their understanding of model behaviour. Other work has highlighted the influence of folk concepts of behaviour~\cite{jacovi2023diagnosing,malle1997folk}, bounded rationality in practitioners' interpretation of XAI~\cite{kaur2024interpretability} and cognitive biases relevant to the design of XAI systems~\cite{wang2019designing}. Therefore users' cognitive strategies and prior beliefs shape what they derive from explanations~\cite{chen2022machine}.

However, misunderstanding cannot be attributed to user cognition alone. Explanations are selective representations that may omit properties necessary for their correct interpretation. %
\citet{bove2024explanations} distinguished system-side from user-side explanation failures, while \citet{jacovi2023diagnosing} identified missing representational and causal components of explanatory narratives as potential sources of misunderstanding. These works broaden the account beyond user error, but they primarily characterise possible failure conditions in conceptual terms rather than tracing how they affect users' reasoning in practice.

What remains less understood is how explanation-side omissions and user-side reasoning interact to produce misunderstanding. In particular, when users seek information that an explanation does not provide, do they recognise that the information is unavailable, or rather confabulate assumptions that allow them to infer it anyway? Our work examines this process empirically by tracing how users reason about information that falls outside of an explanation's scope. %

\subsection{Interactive and Conversational XAI}

To support more accurate understanding of AI models, XAI has increasingly moved from one-shot explanations towards interactive forms of explainability. This shift reflects the view of an explanation as an iterative and social process involving clarification and follow-up questions~\cite{miller2019explanation,madumal2019grounded}. Interactive XAI systems allow users to explore, select and compare explanations rather than passively receiving a fixed explanation~\cite{bertrand2023selective,bhattacharya2023directive}. Prior work has explored interaction through explanatory debugging~\cite{kulesza2015principles}, interactive exploration of model behaviour~\cite{cheng2019explaining} and interfaces that contextualise or compare explanations~\cite{bove2022contextualization,bove2023investigating}.

More recently, conversational XAI has used natural language interaction to make explanation access more flexible. TalkToModel maps users' questions to model analysis and explanation operations~\cite{slack2023explaining}, whereas InterroLang supports conversational explanations for language models~\cite{feldhus2023interrolang}. Other systems connect conversational interfaces with external XAI tools to support a broader range of explanation requests~\cite{wang2024llmcheckup,wang2024coxql,vanbrabant2025echo}. These conversational systems can adapt explanations to the evolving information needs of their users and reduce the burden of knowing which explainability method to use~\cite{mindlin2024measuring,mindlin2025dialogue,zhang2025conversational}. %

Evaluation of these systems has primarily focused on whether the interaction pattern improves users' understanding of the underlying AI system~\cite{bertrand2023selective,cheng2019explaining,slack2023explaining,mindlin2024measuring,zhang2025conversational}. Less attention has been paid to whether users come away misunderstanding the explanations themselves or whether conversational interactions can detect and rectify such misunderstandings once they arise. Our work addresses this gap and further refines the conversational explainability paradigm to help users recognise unfounded interpretations and revise their understanding of an explanation.

\section{AI Explanations and Conversational Explainer}

In this section, we describe the study context used to examine explanation misinterpretation. We then introduce an initial conversational explainer, based on existing conversational XAI interaction patterns, to investigate whether this form of interaction can help users revise out-of-scope interpretations.

\begin{figure}[t]
    \centering
    \begin{minipage}[c]{0.48\linewidth}
        \centering
        \begin{subfigure}{\linewidth}
            \centering
            \includegraphics[width=\linewidth]{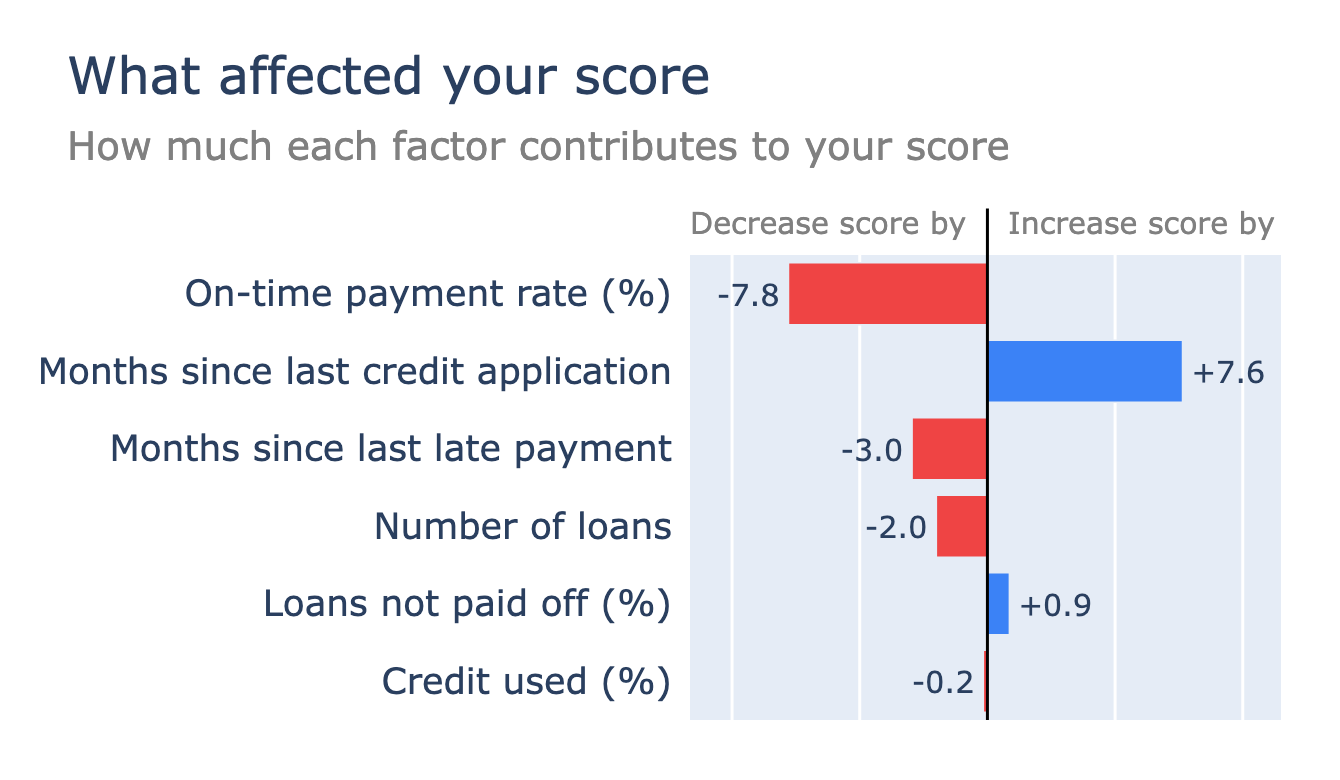}
            \caption{Local feature importance.}
        \end{subfigure}

        \vspace{0.5em}

        \begin{subfigure}{\linewidth}
            \centering
            \includegraphics[width=\linewidth]{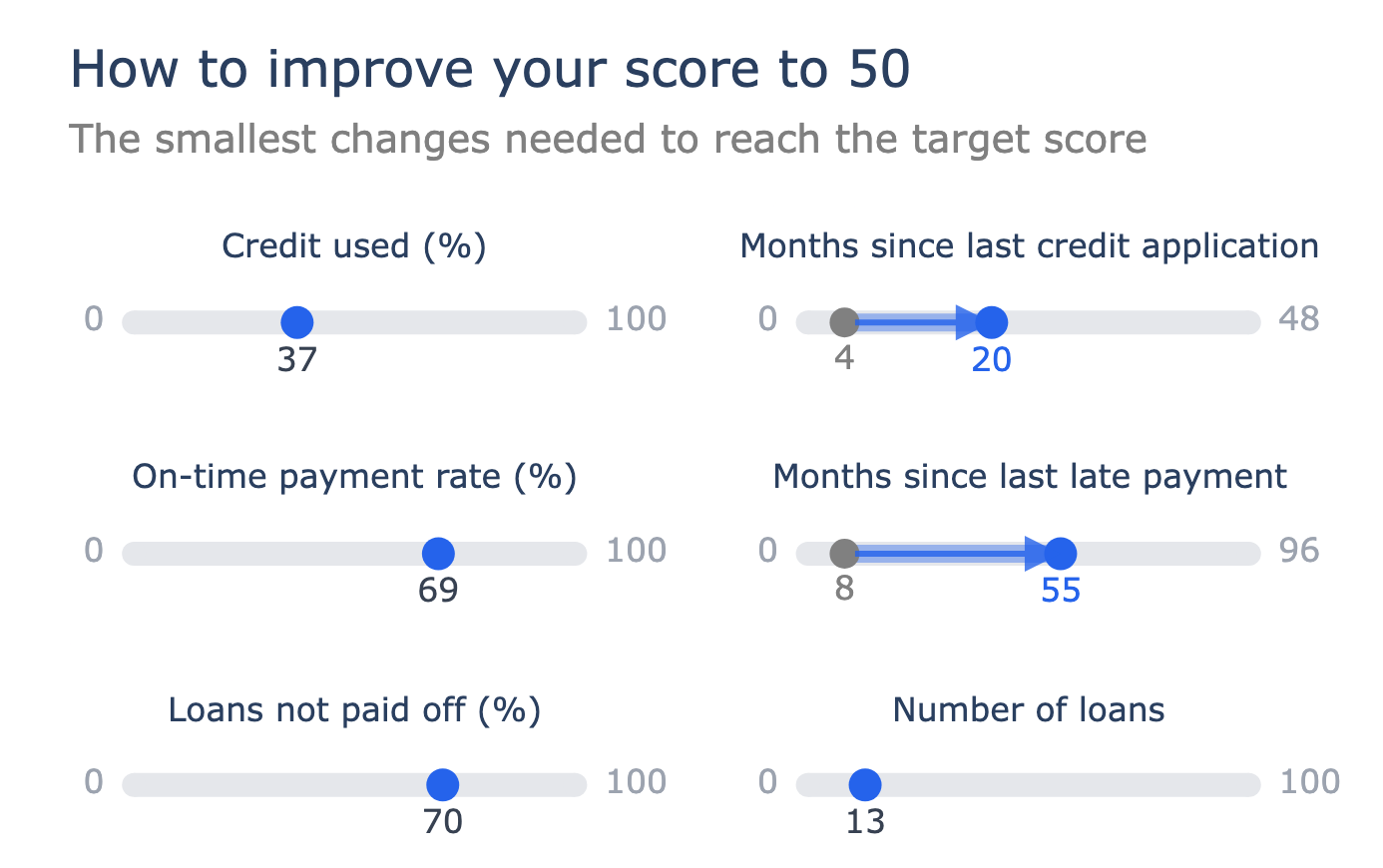}
            \caption{Counterfactual explanation.}
        \end{subfigure}
    \end{minipage}
    \hfill
    \begin{minipage}[c]{0.48\linewidth}
        \centering
        \begin{subfigure}{\linewidth}
            \centering
            \includegraphics[width=\linewidth]{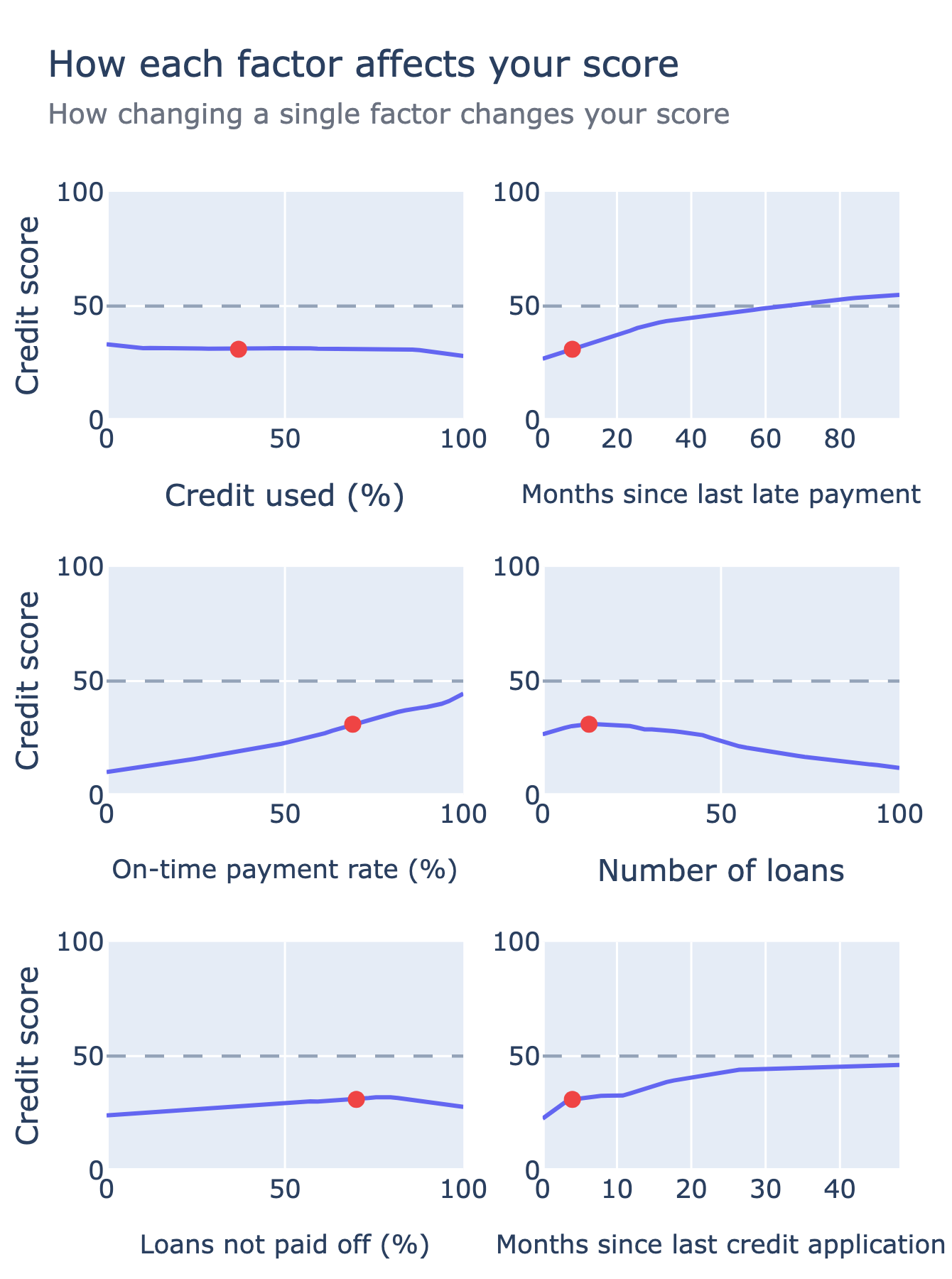}
            \caption{Individual conditional expectation.}
        \end{subfigure}
    \end{minipage}
    \caption{Three local explanations generated for a sample applicant.}
    \label{fig:three-explanations}
\end{figure}

\subsection{Scenario, AI Model and Explanations}\label{sec:context}

To examine how users interpret AI explanations and where misunderstandings arise, we used a common decision scenario and a fixed set of explanation methods across both the qualitative and quantitative phases of our study. Specifically, we used the Home Equity Line of Credit (HELOC) dataset, a common benchmark in XAI research~\cite{fico2018explainable}. The original task predicts the risk of an applicant defaulting on a line of credit. Because this scenario was less familiar to participants in our study setting, we re-contextualised it as a credit card limit increase application. The AI model produced a score for each applicant representing the likelihood of approval. This framing made the task more relatable to the participants while preserving the structure of the original prediction problem.

We trained a three-layer neural network as the underlying opaque AI model. From the 23 features in the HELOC dataset, we ranked features by information gain with respect to the target variable~\cite{louppe2013understanding} and used the six highest-ranked features to train the model, reducing the amount of information participants needed to consider~\cite{miller1956magical}. The description of the six features can be found in Figure~\ref{fig:survey-intro}. The resulting model achieved 73\% accuracy.

Because our study concerns lay decision subjects interpreting predictions about individual applicants, we focused on local explanations. We selected three commonly used methods that provide complementary information: local feature importance, counterfactual explanations and individual conditional expectation. Local feature importance was generated using SHAP~\cite{lundberg2017unified}, counterfactuals were based on growing spheres~\cite{laugel2017inverse} and presented with an established visualisation~\cite{gomez2020vice}, and ICE explanations were computed following the standard formulation~\cite{goldstein2015peeking}. Figure~\ref{fig:three-explanations} shows the three explanations generated for an applicant in the test dataset. We used established visualisation formats for all the explanations from prior XAI research rather than introducing novel presentation designs to maximise their comprehensibility. 

\begin{figure}[t]
    \centering
    \includegraphics[width=0.5\linewidth]{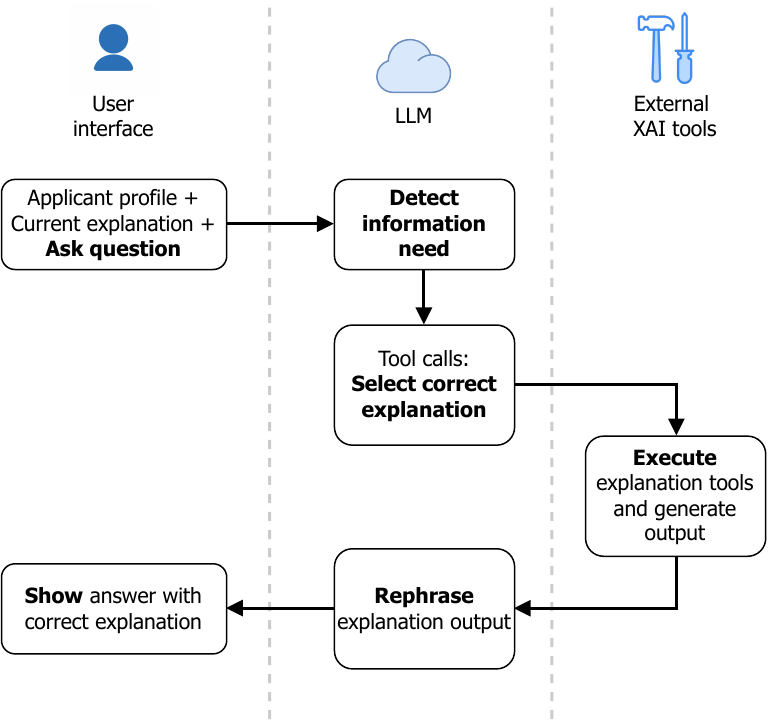}
    \caption{The architecture of our initial conversational explainer.}
    \label{fig:initial-chatbot}
\end{figure}

\subsection{Initial Conversational Explainer}

Conversational XAI allows users to express information needs in natural language and receive relevant AI explanations in response~\cite{slack2023explaining,feldhus2023interrolang,wang2024llmcheckup,mindlin2024measuring}. This makes it a natural starting point for supporting users who may need information beyond what a single explanation can provide. We therefore first implemented an \emph{initial conversational explainer} following the state-of-the-art interaction patterns from conversational XAI, where users' questions are mapped to appropriate explainability methods. The system interprets a user's question, identifies the underlying information needs and selects an explanation capable of addressing them. For example, a question about which features contribute the most to a particular prediction is addressed using local feature importance.

We implemented the system using a tool-augmented large language model (LLM) architecture. Rather than relying on the LLM to generate explanation values directly~\cite{he2025conversational} or using a separate fine-tuned intent-mapping model~\cite{slack2023explaining,mindlin2025dialogue}, we use the LLM to interpret natural language questions and select from a predefined set of explanation operations~\cite{vanbrabant2025echo}. The selected explanation method is then executed programmatically on the underlying AI model and (training) data. This separation preserves the LLM's natural language understanding ability while grounding explanatory outputs in actual XAI methods, reducing the risk of fabricated explanations. Figure~\ref{fig:initial-chatbot} illustrates the architecture of our initial conversational explainer.

This initial system provides a useful starting point for examining whether conversational access to appropriate explanatory information is sufficient to address explanation misunderstanding. In the following interview study we first investigate how misunderstandings arise and then use the initial conversational explainer to help users revise those misunderstandings. Where this interaction proves insufficient, the observed failures can provide the basis for refining the conversational paradigm.

\section{Interview Study}

To investigate how users reason when they seek information beyond the scope of an explanation, we conducted a semi-structured interview study examining the participants' interpretations and reasoning processes. 
The study also served as a formative evaluation of our initial conversational explainer, allowing us to examine whether explanation selection alone was sufficient to rectify misunderstanding and to identify how the interaction should be refined. The study was approved by our institutional Human Ethics Advisory Committee.

\subsection{Study Procedure}\label{sec:interview-procedure}

The interview began by introducing participants to the decision scenario and AI model described in Section~\ref{sec:context}. It then proceeded in three stages. Stage~1 examined how misunderstandings emerged when participants interpreted individual explanations. Stages~2 and~3 then examined whether access to additional explanations, through a dashboard or conversational interaction, could rectify those existing misunderstandings.

\subsubsection{Stage 1: Interpreting Individual Explanations}\label{sec:interview-stage1}

Participants first reviewed the profile of a sample applicant shown in Figure~\ref{fig:alex-profile}. For ease of reference, we refer to this applicant as Alex. Then, they examined the three explanations for this applicant, which are shown in Figure~\ref{fig:three-explanations}. The explanations were presented one at a time. For each explanation, the participants were given two minutes to inspect it and were then asked to answer four questions using only the information available in that explanation:
\begin{description}
\item [Feature (importance) ranking:] Which three features had the biggest impact on Alex receiving a score of 31?
\item [Feature effect:] Did Alex's current value of feature X contribute positively or negatively to the score he received?
\item [Feature direction:] If Alex's value of feature X increased while the other factors stayed the same, would the AI system predict a higher or lower score for him?
\item [Action:] What could Alex change for the AI system to predict a score of 50 for him?
\end{description}
After each answer, the participants were asked to explain how they arrived at the conclusion. This procedure was repeated for all three explanation methods. 

These questions represent distinct information needs that users may expect to satisfy with an explanation~\cite{liao2020questioning}. Crucially, no single explanation could correctly answer all four. Therefore, our four questions included both \emph{within-scope} questions for which the requested information was directly supported by the explanation as well as \emph{out-of-scope} questions for which it could not be inferred from the explanation alone. The participants' verbal reasoning allowed us to examine whether they recognised these information gaps or instead confabulated their own assumptions about the explanation, using it to construct an answer.

\subsubsection{Stage 2: Rectify Misunderstanding through Explanation Dashboard}

Subsequently, the participants were shown all three explanations together in a dashboard. Because the explanations provided complementary information, the dashboard collectively contained the information needed to answer all four questions. Participants revisited each question and were asked to identify which explanation they used to obtain the supporting information while also explaining how they arrived at their answer.

This stage examined whether out-of-scope misunderstanding was primarily caused by the limited information available in a single explanation. If so, providing multiple complementary explanations should allow participants to locate the missing information and revise their earlier answers. If, however, misunderstanding reflected an incorrect interpretation of the original explanation, simply making additional information available might be insufficient. We therefore examined whether participants shifted their focus to the explanation appropriate for each information need or continued to rely on their earlier interpretations despite the relevant information being available elsewhere in the dashboard.

Participants were also asked to predict the AI score for a second applicant using only the explanations shown for Alex. This transfer task examined whether they recognised that the explanations were specific to Alex or instead generalised them to another applicant.

\subsubsection{Stage 3: Rectify Misunderstanding through Conversational Explainer}

Next, participants interacted with our initial conversational explainer to revisit their understanding of one explanation. They first selected an explanation they wanted to examine further, then asked the system questions about the AI model or its decisions. In response, the conversational explainer provided additional AI explanations intended to address those questions. After this interaction, participants were asked again whether their selected explanation itself could answer each of the four questions.

This stage examined whether routing users to an explanation that addressed their question is sufficient to rectify an existing misunderstanding of the original explanation. Notably, although the system could provide relevant explanatory information, it did not explicitly indicate when or why the user's earlier inference was unsupported by the original explanation, 
nor did it clarify why the newly provided explanation was more appropriate for that information need.

\subsection{Participants}

We recruited nine participants through flyers distributed on social media. Participants were required to be at least 18 years old, have no formal training in machine learning and have previously applied for a credit card. These criteria were intended to recruit lay participants who were familiar with the use case scenario but lacked specialist AI knowledge. Interviews were conducted in person or via videoconferencing according to participant preference. Each session lasted approximately 70--90 minutes and the participants received a \$30 gift card. We concluded the recruitment after nine interviews because the participants' responses had become largely convergent and later interviews did not introduce substantively new patterns relevant to our research questions.

\subsection{Qualitative Analysis}
We analysed the interview data using thematic analysis following the method described by \citet{braun2012thematic}. We first reviewed the interview transcripts and generated an initial set of codes capturing participants' interpretations, assumptions and reasoning processes. We then iteratively grouped related codes into candidate themes and refined them. The final themes were selected to capture recurring patterns relevant to our research questions. In reporting the results, the participants are referred to as P1--P9 to preserve anonymity. Only necessary grammatical corrections were made to the participants' quotes when reporting the results.

\section{Interview Study Results}

Across the three stages, we observed a recurring pattern. Participants often did not recognise that an explanation provides only a partial view of an AI model and therefore cannot support every information need. They correctly interpreted information explicitly represented by an explanation while still drawing unsupported conclusions when their information needs extended beyond its scope.

We attribute this process to \textbf{leaky explanations}, drawing a parallel to \emph{leaky abstractions} in software development~\cite{spolsky2002law}. %
AI explanations abstract away details of their construction as well as many underlying (modelling) assumptions. However, these hidden properties determine the explanatory scope of these insights, i.e., what information can and cannot be validly inferred from them. An explanation thus becomes leaky when users require knowledge of these omitted properties to correctly interpret the explanation. 
Because this scope is not made available, users substitute their own (confabulated) assumptions about how the explanation works when seeking information beyond its scope. 
When those assumptions diverge from the explanation's underlying mechanism, the resulting interpretation can seem plausible but remains unsupported by the explanation.

\begin{table}[t]
\caption{Our participants' information needs and the assumptions they used to interpret each XAI method for each information need. The \metneed{} symbol indicates that the information need could be supported by the explanation, whereas the \unmetneed{} symbol indicates that it could not be correctly inferred from the explanation alone. The description of each information need question is provided in Section~\ref{sec:interview-stage1}.}
\label{tab:information-needs}
\centering
\small
\begin{tabularx}{\linewidth}{L{0.13\linewidth} L{0.2\linewidth} L{0.10\linewidth} Y}
\toprule
\textbf{XAI method} & \textbf{Information need} & \textbf{Participant} & \textbf{Assumption} \\
\midrule
\multirow{8}{=}{Local feature importance} & \metneed{Feature ranking} & All & Numerical values represent feature importance \\ \cmidrule{2-4}
    & \metneed{Feature effect} & All & Negative contributions represent a negative impact on the current prediction \\ \cmidrule{2-4}
    & \unmetneed{Feature direction} & P1, P2, P4, P9 & The magnitude and sign of a feature's contribution indicate how changing that feature would affect the prediction \\ \cmidrule{2-4}
    & \unmetneed{Action} & P1, P4, P6 & The directions of feature contributions indicate how the features should be changed to improve the outcome \\
\midrule
\multirow{9}{=}{Counterfactual explanation} & \unmetneed{Feature ranking} & P2, P3, P4, P5, P6, P7, P9 & Factors required in the smallest set of changes are considered more important to the current prediction \\ \cmidrule{2-4}
    & \unmetneed{Feature effect} & P5, P6, P8, P9 & The direction in which a feature must change in the counterfactual indicates how its current value affected the current prediction  \\ \cmidrule{2-4} %
    & \unmetneed{Feature direction} & P6 & The effect indicated by the counterfactual continues in the same direction as the feature value changes further \\\cmidrule{2-4}
    & \metneed{Action} & All & The suggested feature changes must be made together to achieve the counterfactual outcome \\
\midrule
\multirow{13}{=}{ICE} & \unmetneed{Feature ranking} & All & The magnitude or shape of the curve reflects the relative importance of the feature to the current prediction
\\ \cmidrule{2-4}
    & \unmetneed{Feature effect} & P1, P2, P4, P5, P7, P8 & The position of the curve at the current feature value indicates the feature's contribution to the current prediction \\ \cmidrule{2-4}
    & \metneed{Feature direction} & All & The direction of change in the curve indicates whether increasing or decreasing the feature value will improve the prediction \\ \cmidrule{2-4}
    & \metneed{Actions} & All & The effects of changing multiple features can be combined and accumulated to determine the alternative prediction \\ \cmidrule{2-4}
    & \unmetneed{Generalisability to other instances} & All & Trends and slopes for one applicant also apply to other applicants \\
\bottomrule
\end{tabularx}
\end{table}

\subsection{Finding 1: Leaky Explanations Led Users to Fill in the Left-out Explanatory Scope with Their Own Assumptions}%

In Stage~1, the participants generally interpreted information correctly when it was directly supported by an explanation. However, when they sought information beyond the explanation's scope, they rarely concluded that it was unavailable. Instead, they introduced assumptions about how the explanation worked and used these assumptions to derive the information they wanted. Table~\ref{tab:information-needs} summarises these information needs and assumptions.

For local feature importance, the participants often interpreted feature ranking and feature effect correctly and these pieces of information were directly represented. However, P1, P2, P4 and P9 also treated the sign and magnitude of a feature's current contribution as evidence of the direction in which this feature should be modified. For example, P1 reasoned that ``those three features contributed negatively, so their values were low and we need to increase them''. 
This inference is not supported by this explanation alone: knowing that a current feature value contributed negatively does not establish how the prediction will change as that feature value increases.

Counterfactual explanations showed a similar form of leakiness. Although counterfactuals directly identify the smallest changes that could produce a desired outcome, seven participants treated features appearing in the smallest counterfactual change as the most important features for the original prediction. Four participants also inferred feature effect from the direction of the counterfactual change. For example, P6 reasoned that ``if one [feature] needs to [be] increase[d] in the smallest changes, it must have contributed negatively to the result before''.

ICE explanations produced the same pattern. Although ICE directly shows how a prediction will change as a feature varies for a particular instance, all the participants attempted to infer feature ranking from properties such as the slope or shape of the curves. As P2 explained: ``on-time payment rate has the steepest curve so it affected the current result the most''. 
Participants also generalised these relationships across applicants, treating trends observed for one instance as relationships that would hold for others because they assumed the AI model applied the same feature-response patterns across instances.

Across the explanation types, the participants responded to out-of-scope information needs by confabulating plausible assumptions about properties that the explanations did not represent. This is when the explanations became leaky: determining whether the desired inference was valid required knowledge of the properties hidden by the explanations' abstraction. Because the participants did not have access to those properties, they substituted their own assumptions about how the explanations worked. When these assumptions diverged from the explanations' actual mechanism, they produced seemingly plausible misinterpretations.

\subsection{Finding 2: Multi-Explanation Dashboard Cannot Resolve Out-of-Scope Misinterpretations}

In Stage~2, providing all three explanations made the relevant information available for each question, but this did not allow the participants to reliably revise their earlier interpretations. The participants rarely treated the dashboard as a set of explanations with distinct information scopes. Instead, they often \textbf{anchored on one explanation} they found particularly intuitive or informative and continued to use it across multiple information needs. Other explanations were then used either to confirm this interpretation when they appeared consistent with it or disregarded when they were difficult to reconcile.

For example, P1 inferred feature ranking from ICE and used local feature importance to confirm this conclusion. Rather than recognising that the two explanations supported different types of information, P1 treated their apparent agreement as evidence that their original interpretation of ICE was valid. Similarly, P6 inferred a feature's effect from ICE and used the absence of that feature in a counterfactual explanation to reinforce the same conclusion: ``the number of loans had a positive effect on the current prediction since it is at the peak of the curve and also it is not needed to increase in the smallest changes''. In both cases, access to an additional explanation reinforced rather than corrected the participants' original interpretation of ICE.

Other participants disregarded explanations that conflicted with the interpretations they had already formed. P7, for example, stated regarding local feature importance that ``it doesn't make sense to me what the decrease and increase are for'' and instead relied only on ICE to infer both feature effects and feature importance, even though ICE did not directly support these conclusions.

These findings show that out-of-scope interpretations of an explanation cannot be resolved simply by making the missing information available elsewhere. The participants were likely to continue to treat an out-of-scope inference as valid and interpret other explanations while being biased by existing beliefs. In some cases, additional explanations even reinforced the original interpretations when the participants selectively treated apparent agreement as confirmation -- a phenomenon consistent with the confirmation bias~\cite{kliegr2021review}. 
Thus, rectifying out-of-scope interpretations arising from leaky explanations requires the users to recognise that the inferences they are making exceed the scope of the explanations they rely on.

\subsection{Finding 3: Conversational Explainer Cannot Rectify Existing Misinterpretation}\label{sec:finding-3}

In Stage~3, the conversational interaction made it easier for the participants to obtain the relevant explanatory information, but existing misunderstandings often persisted. When the system answered a user question with another explanation, it did not explicitly tell them why the original explanation could not answer their question in the first place. The participants therefore could not revise their understanding of what the original explanation was able to communicate. %

For example, while examining local feature importance, P1 viewed global feature importance and questioned why the feature rankings differed: ``I don't know why this [global feature importance] shows a different ranking than the local one; this is very confusing''. Although the global explanation provided information about global importance, it did not explicitly clarify why local feature importance could not be used to infer global importance.

Similarly, while exploring a counterfactual explanation, P8 continued to regard the features changed in a counterfactual as ``important'' ones even after reading about local feature importance because they still assumed that the features appearing therein signified their importance. %
Likewise, P7 questioned why feature ranking cannot be obtained from ICE after reading about local feature importance and insisted that ``I would rather use ICE to find the information since it not just tells me where I am now but also [about] the changes''. In both cases, receiving the appropriate information did not challenge the assumption underlying the original misinterpretation.

These cases suggest that providing the correct explanation and rectifying misinterpretations of a leaky explanation are distinct processes. When out-of-scope information need is grounded in an unsupported assumption of the current explanation, simply providing another explanation does not necessarily reveal what was wrong with the original inference. 
Rectifying such mistakes therefore requires making the scope mismatch explicit before directing the users to an explanation whose scope can provide the information they seek.

\section{Explanation Navigator}

Our interview findings reveal a central challenge posed by leaky explanations: the scope of an explanation constrains what its users can validly infer, yet this scope is often not visible in the explanation itself. As a result, the users may cross that boundary without recognising it and confabulate assumptions to derive information the explanation does not provide. Simply supplying additional explanations does not reliably correct these interpretations, as the users remain unaware that their original inference is unsupported. Rectifying out-of-scope interpretations therefore requires helping the users recognise when and why their inference exceeds the explanation's scope. Based on this insight, we identify three design considerations to refine our initial conversational explainer, implementing them in a framework called \emph{Explanation Navigator}. %

\subsection{Design Considerations}\label{sec:design-considerations}

Misinterpretations of AI explanations are similar to misconception problems studied in knowledge-revision research. Once learners form a misconception, simply presenting correct information does not necessarily replace it; the prior conception can persist unless the conflict between the existing and new information is made explicit~\cite{prinz2021counteracting}. The \emph{knowledge revision components} framework similarly argues that previously encoded knowledge cannot simply be erased and replaced~\cite{kendeou2014krec}. Revision is more effective when the existing conception and a plausible alternative are activated together, allowing new information to update the prior representation~\cite{kendeou2019knowledge}.

Research on refutation texts operationalises a similar principle by explicitly identifying an incorrect conception, explaining why it is incorrect and providing a correct alternative. Such interventions can facilitate a conceptual change more effectively than presenting correct information alone~\cite{schroeder2022refutation}. Furthermore, \citet{rich2017belief} found that misconceptions are revised more successfully when the refutation and supporting explanation are presented together rather than merely stating that the original belief is incorrect. Therefore, both refutation and alternative answers are essential to revising misconceptions.

Drawing on these principles and our interview findings, we derive three design considerations (DCs) to refine our initial conversational explainer.

\paragraph{\normalfont\bfseries{}DC1: Detect Mismatches between Information Needs and Explanation Scope}
Our first finding showed that users often did not recognise when the information they sought exceeded what the current explanation could support. %
A conversational system should therefore treat the user's emerging information need as a signal to check whether the requested conclusion falls within the scope of the explanation currently being interpreted. Detecting this mismatch is necessary before a correction can occur because users may otherwise continue to treat the current explanation as valid evidence for an unsupported inference.

\paragraph{\normalfont\bfseries{}DC2: Make the Relevant Explanation Scope Explicit}

Our findings showed that even when the appropriate information was available elsewhere, participants often retained their original interpretation because they were never told why the original explanation could not support their inference. Therefore, after detecting a mismatch, the conversational system should make it explicit by communicating both the explanation's scope and its property that gives rise to it. 
This approach is consistent with knowledge-revision research, which emphasises making conflicts between prior and alternative interpretations explicit~\cite{kendeou2014krec,kendeou2019knowledge}. For example, when users attempt to infer feature importance from ICE, the system can clarify that ICE is generated by varying one feature across hypothetical values while treating the remaining features as fixed. It therefore shows how the prediction responds to changes in that feature but does not quantify how much the feature contributed to the current prediction.

Notably, the scope disclosure should be responsive to the user's information need rather than presenting the full scope or technical construction details of an explanation upfront. Explainability methods may involve many modelling assumptions and boundaries; exposing them all would require the users to determine their relative relevance, which can be challenging and add cognitive load. The system should instead disclose the relevant scope as well as the property necessary to explain why a particular inference is unfounded. %

\paragraph{\normalfont\bfseries{}DC3: Resolve Unmet Information Needs with Complementary Explanations}

Making the scope explicit corrects the interpretation of the current explanation but leaves the user's original information need unresolved. Refutation-text research highlights the value of providing a coherent alternative rather than merely rejecting an existing conception~\cite{schroeder2022refutation,rich2017belief}. The conversational explainer should therefore follow the scope disclosure with a complementary explanation whose scope can address the information sought by the users. %
For example, after clarifying why a counterfactual explanation cannot establish feature importance, the system can provide a feature importance explanation and explain why it is appropriate for that information need.

DC2 and DC3 serve complementary purposes in repairing misinterpretation: DC2 addresses why the current explanation cannot answer the question, while DC3 addresses where the requested information can instead be obtained.  %

\begin{figure}[t]
    \centering
    \includegraphics[width=0.8\linewidth]{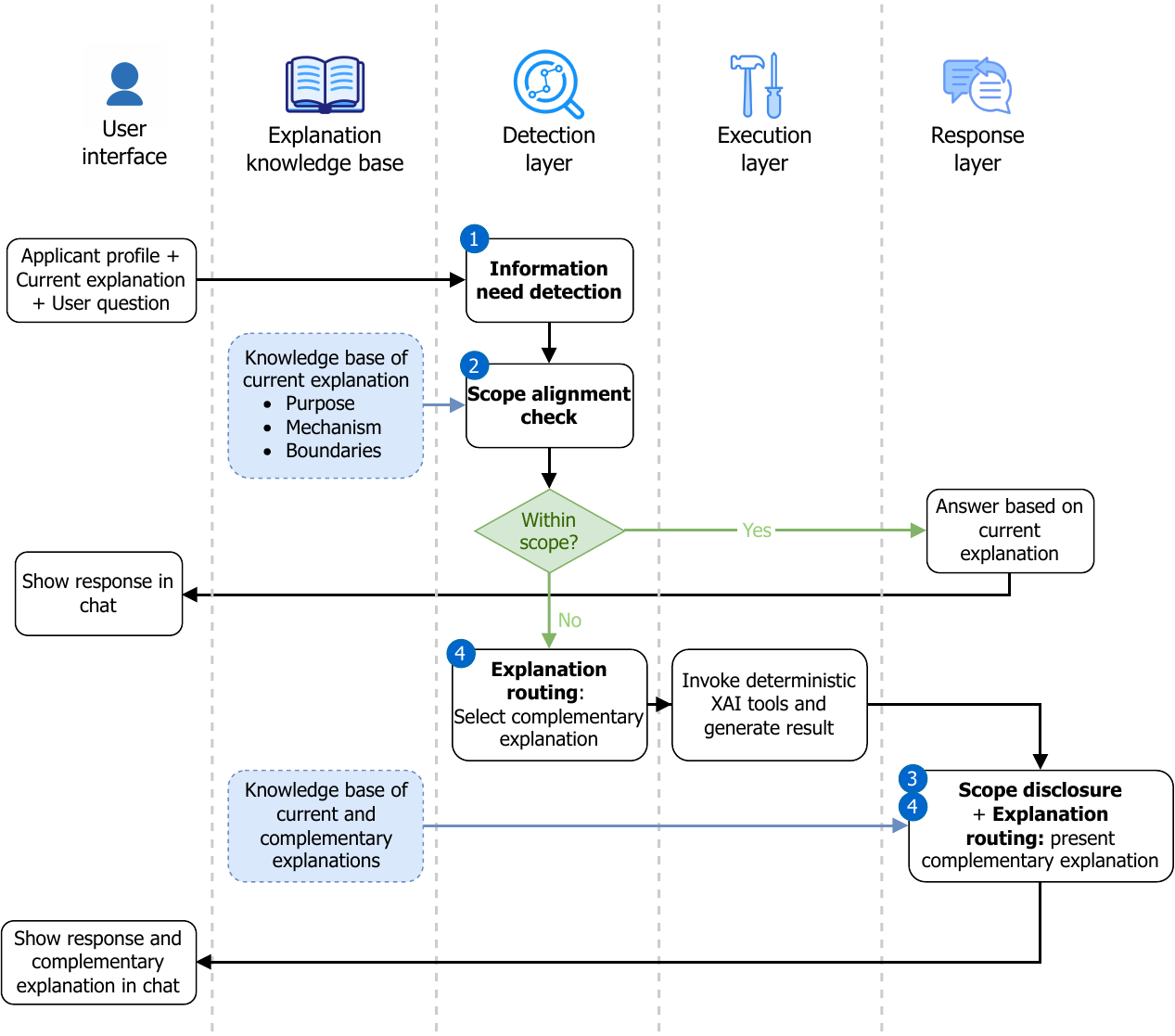}
    \caption{Architecture of the explanation navigator; the four-step workflow is implemented using four distinct components.}%
    \label{fig:architecture}
\end{figure}

\subsection{Explanation Navigator Workflow}\label{sec:exp-navigation}

Building on our initial conversational explainer, we implement the design considerations within a new interaction framework called the \textbf{Explanation Navigator}. %
Unlike existing conversational XAI patterns that directly route each information need to an appropriate explanation, our Explanation Navigator first assesses whether the current explanation can support the requested information. When it detects a mismatch, it explicitly communicates the relevant scope before directing the user to a complementary explanation. 

The interaction with the users begins by presenting them with a single AI explanation, after which the users can express additional information needs through natural language questions. For each user question, the Explanation Navigator performs four steps:
\begin{enumerate}
    \item \textbf{Information need detection}: It identifies what the user is trying to learn about the AI model from their question. 
    \item \textbf{Scope alignment check}: It assesses whether the current explanation can support this information need by comparing it with the explanation's purpose, mechanism and boundaries.
    \item \textbf{Scope disclosure}: When a mismatch is detected, it explains why the requested inference cannot be supported by the current explanation and makes explicit the relevant scope and underlying explanation properties.
    \item \textbf{Explanation routing}: It identifies and presents a complementary explanation that can support the information need and explains why it is appropriate for the user's question.
\end{enumerate}
The first two steps implement DC1; %
step three realises DC2; %
step four achieves DC3.

When the current explanation already supports the user's information need, the system answers directly using that explanation. The scope disclosure and explanation routing are therefore invoked only when the user's information need exceeds what the current explanation can support.

\subsection{System Architecture}

We implement the Explanation Navigator as a conversational system that extends our initial tool-augmented LLM architecture with the additional ability to reason about the relationship between users' information needs and the scope of the explanation currently being viewed. As shown in Figure~\ref{fig:architecture}, the navigator consists of four components: an explanation knowledge base, a detection layer, an execution layer and a response layer.

\paragraph{\normalfont\bfseries{}Explanation Knowledge Base}
We constructed a knowledge base describing each explanation method supported by the system along three dimensions:
\begin{enumerate}
\item \textbf{Purpose}: The information that the explanation is intended to communicate.
\item \textbf{Mechanism}: A high-level description of how the explanation is generated and the modelling assumptions underlying its construction, which in turn constrain the scope of valid interpretations.
\item \textbf{Boundaries}: The information limits and conclusions that cannot be validly inferred from the explanation, including recurring unsupported interpretations identified in our interview study.
\end{enumerate}

The knowledge base is configured for the model, dataset and explanation methods available in a particular deployment. Examples of knowledge base entries in our implementation are provided in Appendix~\ref{app:knowledge-base}. Explicitly representing this knowledge provides the Explanation Navigator with grounded information for determining whether an explanation can support a user's information need instead of relying on the LLM to infer explanation properties itself.

\paragraph{\normalfont\bfseries{}Detection Layer}
The detection layer uses an LLM with tool-calling capabilities to interpret the user's natural language question. It performs Step~1 from Section~\ref{sec:exp-navigation} to identify the underlying information need as well as Step~2 to compare it with the knowledge base for the explanation that is currently being displayed. If the information need falls within the explanation's scope, the interaction proceeds directly to the response layer. If a mismatch is detected, the layer identifies the relevant scope and explanation properties to disclose and selects a complementary explanation from the supported methods that can address the information need.

\paragraph{\normalfont\bfseries{}Execution Layer}
The execution layer invokes deterministic programmatic functions that run the XAI methods selected by the detection layer on the underlying model and dataset. 
The explanation tools supported in our implementation are summarised by Table~\ref{tab:tool-desc} in Appendix~\ref{app:tech-detail}. The explanation values and visualisations are generated by the corresponding XAI tools rather than by the LLM, making the explanatory artefacts more reliable and verifiable.  %

\paragraph{\normalfont\bfseries{}Response Layer}

The response layer constructs the final conversational response by invoking the LLM with the outputs of the execution layer and the explanation knowledge base. When the current explanation supports the user's information need, the response is grounded based on the explanation itself. When a mismatch is detected, the layer performs Steps~3--4 from Section~\ref{sec:exp-navigation}: it responds by rephrasing the relevant explanation scope and properties in layperson-friendly terms, then presenting the complementary explanation with an appropriate description.

\begin{figure}[t]
    \centering
    \includegraphics[width=\textwidth]{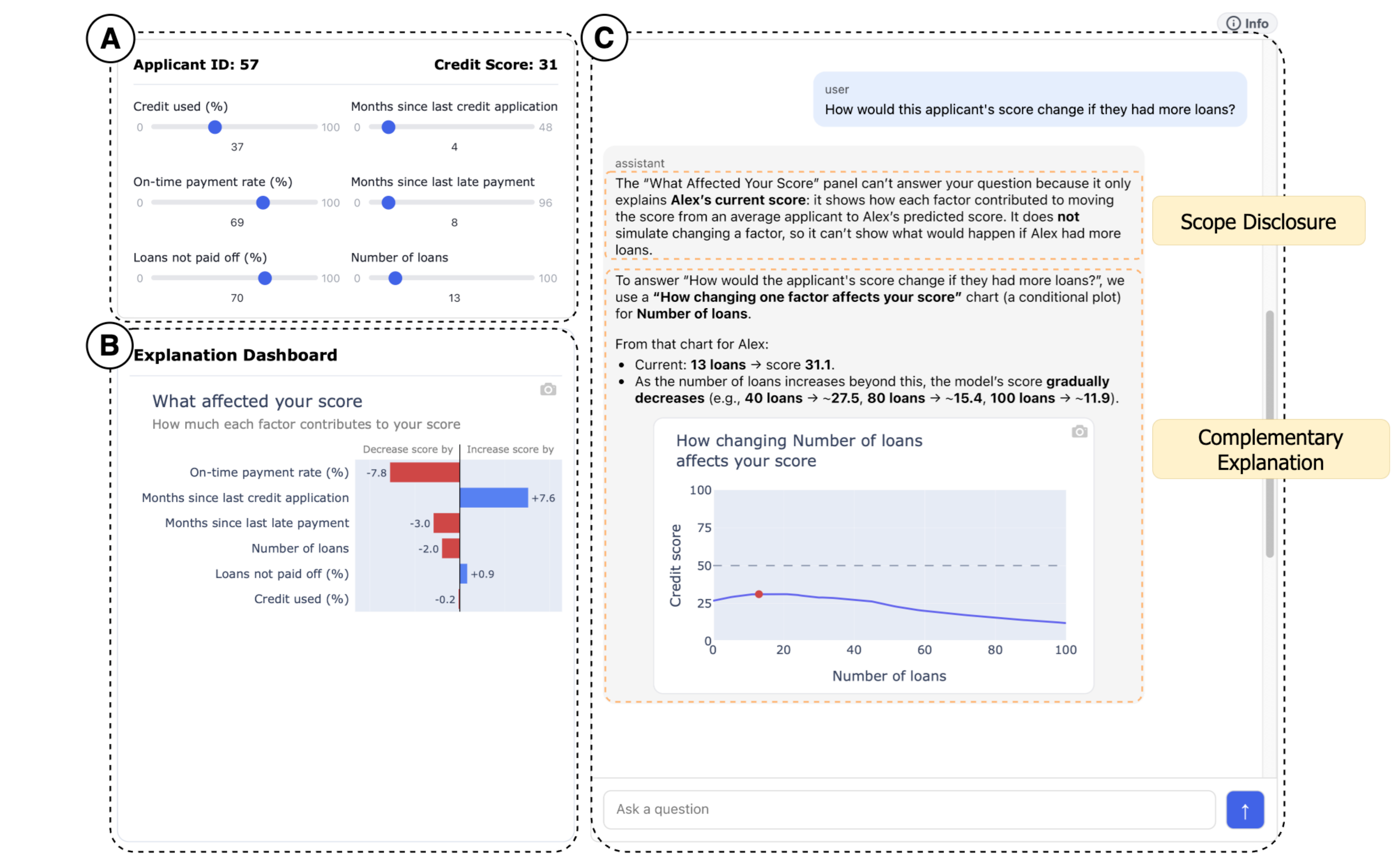}
    \caption{Explanation Navigator interface. Box~(A) displays the applicant profile and AI prediction; Box~(B) presents the explanation currently being interpreted; and Box~(C) supports conversational interactions. In this example, once a mismatch is detected, the system responds by first disclosing the relevant explanation scope, explaining why the current explanation cannot support the requested inference. Then it provides a complementary explanation that addresses the user's information need.}
    \label{fig:interface}
\end{figure}

\subsection{System Interface and Technical Details}

The Explanation Navigator interface contains three main components, which are shown in Figure~\ref{fig:interface}. The profile pane displays one applicant's feature values and predicted score; the explanation pane presents the explanation currently being interpreted; and the conversation pane allows the users to ask questions and displays the navigator's responses, including complementary explanations when needed. Notably, the current explanation remains visible when a complementary explanation is introduced. This preserves the original explanation as interaction context and allows the system to explicitly relate the two explanations and clarify why the original explanation cannot support the inference while the complementary explanation can provide the relevant information. The full technical details of our implementation of the Explanation Navigator are provided in Appendix~\ref{app:tech-detail}. Our code is available at \href{https://github.com/xuanxuanxuan-git/explanation-navigator}{github.com/xuanxuanxuan-git/explanation-navigator}.

\section{Quantitative User Study}

Our Explanation Navigator is designed to rectify the out-of-scope interpretations of leaky explanations identified in our interview study. Next, we conduct an online user study to quantitatively evaluate its effectiveness. Specifically, we examine the two corrective components of our framework -- the scope disclosure and the explanation routing -- to determine whether disclosing an explanation's relevant scope is sufficient to reduce out-of-scope misinterpretations and whether additionally routing users to complementary explanations provides further benefits. The study was approved by our institutional Human Ethics Advisory Committee.

\begin{figure}[t]
    \centering
    \includegraphics[width=0.95\linewidth]{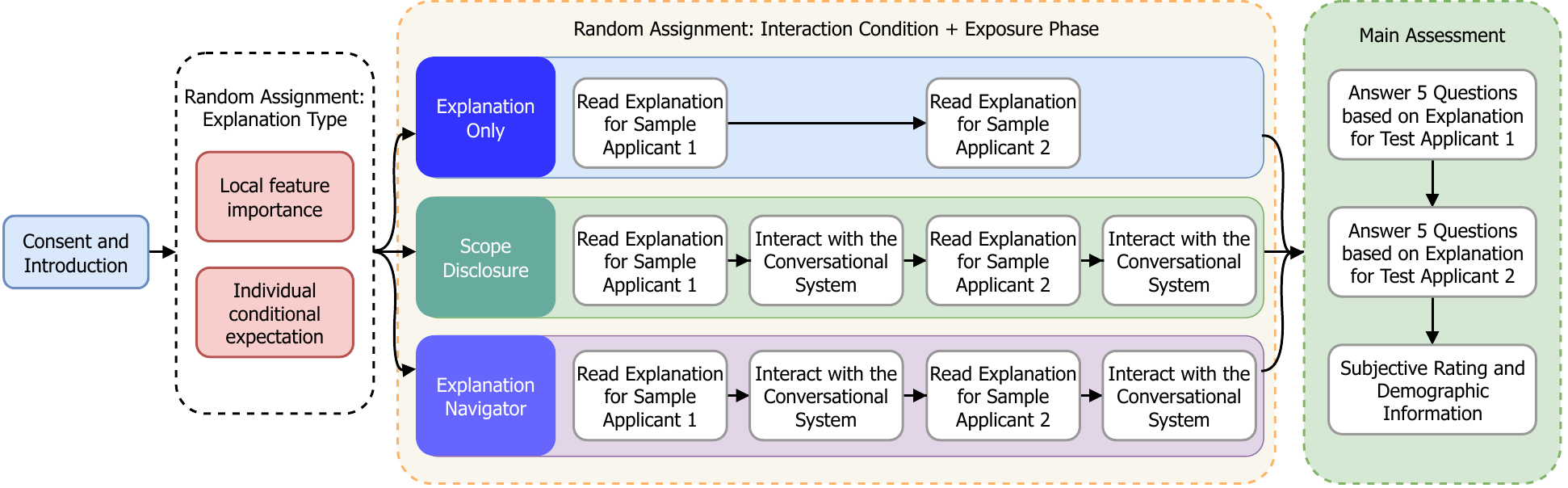}
    \caption{Overview of our online study procedure. The participants were assigned to one of six conditions determined by the explanation type and interaction condition. During the exposure phase, they viewed their assigned explanation for two sample applicants and, where applicable, interacted with the corresponding conversational system. They then completed the main assessment without access to the conversational system. The interfaces presented to the participants in our survey are shown in Figures~\ref{fig:survey-intro}--\ref{fig:llm-interface} in Appendix~\ref{app:survey}.}
    \label{fig:survey-flowchart}
\end{figure}

\subsection{Experimental Design}

We employed a $2 \times 3$ between-subjects design with \textbf{explanation type} and \textbf{interaction condition} as independent variables. The participants were assigned either to a local feature importance explanation or an ICE explanation. We focused on these two explainability methods because our qualitative study revealed especially frequent scope misunderstanding for them.

For each explanation type, the participants were randomly assigned to one of three interaction conditions:
\begin{enumerate}
\item \textbf{Explanation Only (Baseline)}: The participants received the applicant profile and the target explanation together with a brief note describing what the explanation communicates. This represents a common setting in which an explanation is presented without further (interactive) support.

\item \textbf{Scope Disclosure}: The participants interacted with a restricted version of the Explanation Navigator. When the system detected a mismatch between the user's information need and the current explanation, it only communicated the relevant scope and why the requested inference was unsupported, following Steps~1--3 described in Section~\ref{sec:exp-navigation}. It did not perform Step~4 and therefore did not navigate to a complementary explanation to address the users' questions. Example response from this condition is shown in Figure~\ref{fig:scope-disclosure-response} in Appendix~\ref{app:survey}.

\item \textbf{Explanation Navigator}: The participants interacted with the full Explanation Navigator described in Section~\ref{sec:exp-navigation}. When a scope mismatch was detected, the system first disclosed the relevant scope and then provided a complementary explanation capable of addressing the unmet information need.
\end{enumerate}

These conditions were designed to isolate the contribution of the two corrective components offered by our interaction framework. The Scope Disclosure approach tests whether making the relevant explanation scope explicit is sufficient to reduce unsupported inferences. The Explanation Navigator tests the additional benefits of providing a complementary explanation after the scope mismatch has been clarified.

We did not include a separate condition to test our initial conversational explainer, which only routed the users to an explanation matching their question. This interaction pattern had already been examined in our qualitative study (see Section~\ref{sec:finding-3}), where providing the appropriate explanation did not reliably revise the participants' interpretation of the original explanation. Our quantitative study therefore focuses on the corrective mechanisms introduced in response to that finding: whether the scope disclosure is sufficient on its own and whether subsequently providing a complementary explanation further improves understanding.

\subsection{Study Procedure}
The study was hosted on Qualtrics~\cite{qualtrics2026}, with the participants recruited through Prolific~\cite{prolific2026}. Figure~\ref{fig:survey-flowchart} summarises the study procedure.
After providing informed consent, the participants were introduced to the scenario, AI model and predictive task. The context was identical to that used in our interview study described in Section~\ref{sec:context}. 

The participants were then randomly assigned to one of the six experimental conditions. They first completed an exposure phase where they viewed their assigned explanation type for two sample applicants. Those in the Scope Disclosure and Explanation Navigator conditions also interacted with their corresponding conversational systems. For each sample applicant, they were required to ask at least five questions to the system before proceeding, thus encouraging them to explore what information the explanation could and could not support.

All the participants then completed the main comprehension assessment using two previously unseen test applicants. Notably, the conversational system was \emph{not available during the assessment} in any condition. The participants therefore had to answer based on their understanding developed during the exposure phase rather than retrieving the answers directly from the system. This allowed us to evaluate whether the interaction changed the participants' understanding of the explanation itself. The survey interface for all the conditions is shown in Appendix~\ref{app:survey}.

For each test applicant, the participants answered five questions representing different information needs: 
\begin{enumerate}
    \item \textbf{Feature ranking}: Which two factors had the biggest impact on the applicant getting the score of [$\times$]?
    \item \textbf{Feature effect}: Did the applicant's current [feature $\times$] contribute positively or negatively to the current score they received?
    \item \textbf{Feature direction}: If this applicant's [feature $\times$] increased while other factors stayed the same, would the AI system predict a higher or lower score for them?
    \item \textbf{Action}: What (single feature) could this applicant change for the AI system to predict a score of 50 for them?
    \item \textbf{Model-level question}: Is [feature $\times$] the most important factor in the AI system's predictions for all applicants? Or alternatively: Would increasing [feature $\times$] increase the predicted score for other applicants?
\end{enumerate}

Here, feature $\times$ refers to one of the six features used by the AI model and varies across questions. 
The two test applicants were presented in random order and the five questions for each applicant were also randomised, yielding ten assessment questions per participant. Examples of the assessment questions used in the study are shown in Figures~\ref{fig:objective-questions}--\ref{fig:objective-questions-ice} in Appendix~\ref{app:survey}. 
The choices of the questions were informed by the XAI Question Bank~\cite{liao2020questioning}, which has been used in other XAI evaluation studies~\cite{cheng2019explaining,wang2021explanations,slack2023explaining,he2025conversational,xuan2025comprehension}. All the experimental groups answered the same set of questions for their assigned explanation type.

Following the objective assessment, the participants rated the explanations' ease of understanding (``this explanation is easy to understand'') and sufficiency of detail (``this explanation is sufficiently detailed'') on a seven-point Likert scale. These subjective measures are adopted from prior user studies~\cite{lage2019human,bo2024incremental,he2025conversational,hoffman2023measures}. The participants in the two conversational conditions additionally rated whether the system has helped them better understand the explanation (``this chatbot assistant helps me better understand the explanation''). Finally, we collected demographic information and self-reported familiarity with AI and applying for credit.

\subsection{Measures}

Following an assessment approach used in prior work to evaluate users' misinterpretation of AI explanations~\cite{xuan2025comprehension}, our objective assessment included questions that could and could not be answered correctly based on the provided explanations. Each question therefore included a ``cannot tell from this explanation'' response. For each explanation, four of the ten questions were within-scope, meaning that the underlying explanation contained sufficient information to answer them correctly. The accuracy computed for these questions measured \textbf{within-scope comprehension}, that is, whether the participants correctly interpreted the information provided by the explanation.

The remaining six questions were out-of-scope, meaning that the requested information could not be established from the explanation alone. For these questions, the correct response was ``cannot tell from this explanation''. The accuracy computed for these items measured \textbf{out-of-scope recognition}, that is, whether the participants correctly recognised the limits of what could be inferred from the current explanation.

For local feature importance, the within-scope questions are \emph{feature ranking} and \emph{feature effect}. For ICE, the within-scope questions are \emph{feature direction} and \emph{action}. Each information need was assessed for both test applicants. We also report overall accuracy across all ten questions as a summary measure of objective explanation understanding.

As a secondary measure, we collected the participants' subjective rating on a seven-point Likert scale -- coded from \emph{1: strongly disagree} to \emph{7: strongly agree} -- with higher scores indicating more positive evaluation.

Lastly, to analyse the participants' interaction with the conversational systems during the exposure phase, we recorded their interaction time and all the natural language queries they submitted.

\subsection{Participant Recruitment}

To determine the required sample size, we conducted an a priori power analysis using G*Power~\cite{faul2009statistical}. We considered three between-subject conditions for each explanation type, resulting in six conditions overall. Assuming a medium effect size of $f=0.25$, $\alpha=.05$ and power of $1-\beta=.80$, the analysis indicated a requirement of approximately 53 participants per condition, or 318 participants in total.

We recruited 320 participants through Prolific. To promote response quality, the participants were required to have an approval rating of at least 95\%, consistent with prior online XAI studies~\cite{van2021effect,ma2025towards}. Because the study involved inspecting explanation visualisations and interacting with a conversational interface, participation was only allowed using desktop devices. Based on a pilot study, we estimated a study completion time of 15 minutes. The participants were compensated \$6.10, corresponding to an hourly rate of \$24.40.

\section{Online Study Results}

After excluding the participants who failed two attention-check questions, we retained 316 valid responses. For local feature importance, the Baseline, Scope Disclosure and Explanation Navigator conditions included 54, 53 and 51 participants respectively; the corresponding ICE conditions included 54, 52 and 52 participants. The average completion time was 14.3 minutes ($SD=5.6$), with averages of 10.1, 15.7 and 16.2 minutes for the Baseline, Scope Disclosure and Explanation Navigator conditions respectively. 
Overall, 46\% of the participants were aged 30--44 and 50\% were female. 
Most of the study participants reported limited prior AI knowledge: 71\% reported less than moderate familiarity with AI systems and 81\% reported less than moderate familiarity with a credit limit application process. Complete demographic information is provided in Table~\ref{tab:demographics} in Appendix~\ref{app:additional-results}.

\begin{table}[t]
\caption{Generalised Linear Mixed-Effects Model results for explanation understanding accuracy. The explanation type, interaction condition, question category and all interaction terms were included as fixed effects, with the participant and assessment items considered random intercepts. 
Significance indicators: $^*:p<.05$, $^{**}:p<.01$ and $^{***}:p<.001$.}
\label{tab:glmm_anova}
\centering
\small
\begin{tabular}{lrc}
\toprule
\textbf{Fixed Effects \& Interaction Terms} & \textbf{Wald $\chi^2$} & \textbf{$p$}\\
\midrule
Explanation Type & 29.40 & 0.589 \\
Interaction Condition & 255.14 & $< 0.001^{***}$ \\
Question Category & 332.43 & $< 0.001^{***}$ \\
Explanation Type $\times$ Interaction Condition & 2.91 & 0.234 \\
Explanation Type $\times$ Question Category & 0.83 & 0.362 \\
Interaction Condition $\times$ Question Category & 73.82 & $< 0.001^{***}$ \\
Explanation Type $\times$ Interaction Condition $\times$ Question Category & 5.52 & 0.063 \\
\midrule
\multicolumn{3}{l}{Marginal $R^2 = 0.434$} \\
\bottomrule
\end{tabular}
\end{table}

\begin{figure}
    \centering
    \begin{subfigure}[t]{0.4\linewidth}
        \centering
        \includegraphics[width=\linewidth]{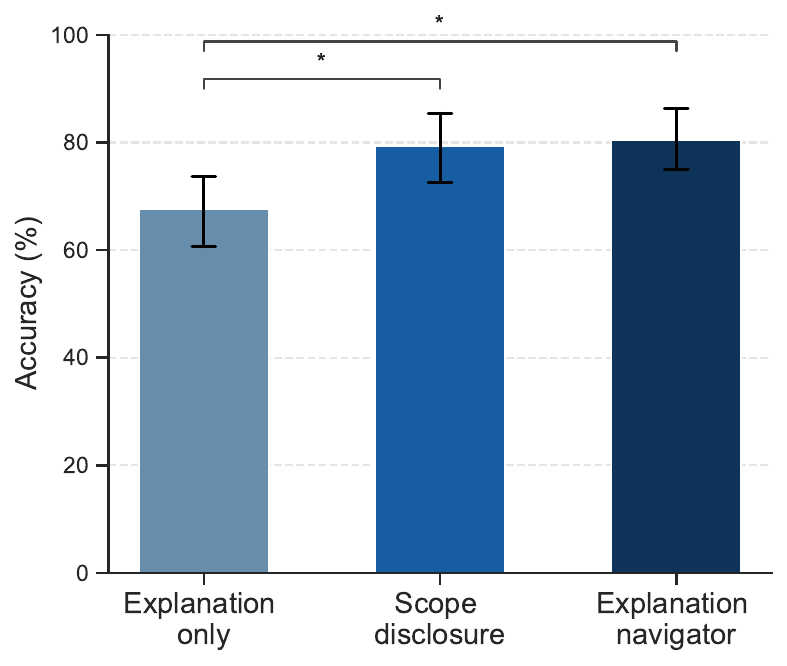}
        \caption{Within-scope comprehension for local feature importance.}
        \label{fig:local-can}
    \end{subfigure}
    \hspace{2em}
    \begin{subfigure}[t]{0.4\linewidth}
        \includegraphics[width=\linewidth]{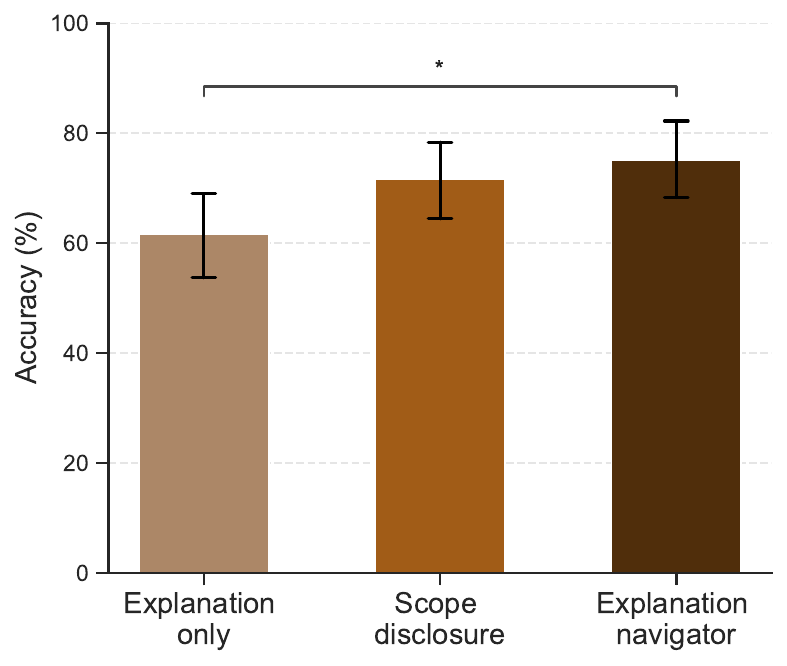}
        \caption{Within-scope comprehension for ICE.}
        \label{fig:ice-can}
    \end{subfigure}
    \\
    \begin{subfigure}[t]{0.4\linewidth}
        \centering
        \includegraphics[width=\linewidth]{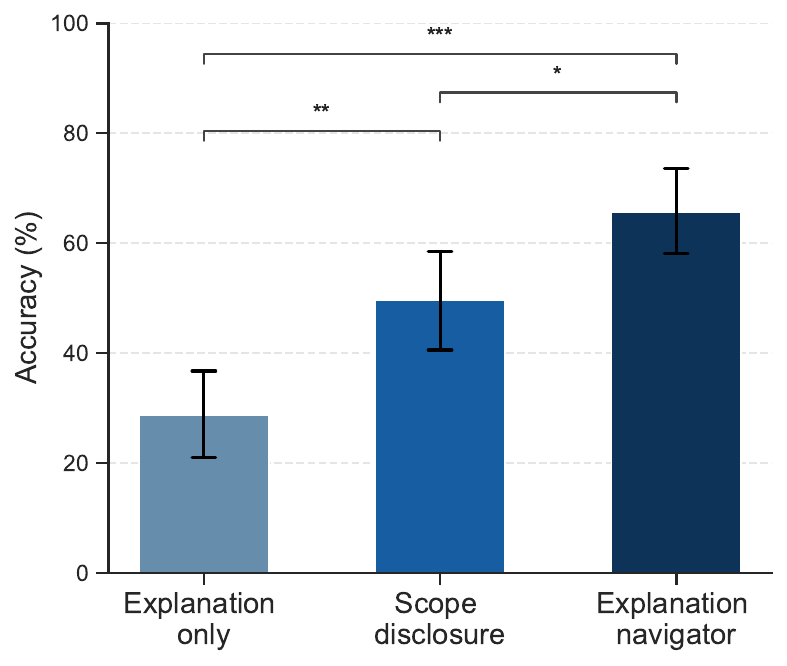}
        \caption{Out-of-scope recognition for local feature importance.}
        \label{fig:local-cannot}
    \end{subfigure}
    \hspace{2em}
    \begin{subfigure}[t]{0.4\linewidth}
        \includegraphics[width=\linewidth]{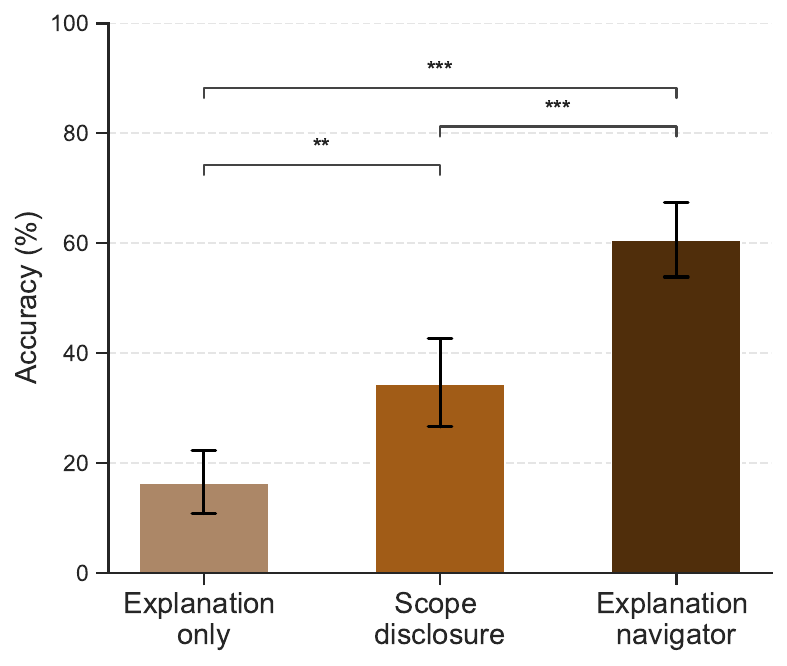}
        \caption{Out-of-scope recognition for ICE.}
        \label{fig:ice-cannot}
    \end{subfigure}
    \caption{Mean comprehension accuracy across our interaction conditions for the local feature importance and ICE explanations. 
    Panels~(\subref{fig:local-can}) and~(\subref{fig:ice-can}) show accuracy for the within-scope questions, whereas Panels~(\subref{fig:local-cannot}) and~(\subref{fig:ice-cannot}) show accuracy for the out-of-scope questions. The error bars indicate 95\% confidence intervals. Significance indicators: $^*:p<.05$, $^{**}:p<.01$ and $^{***}:p<.001$.}
    \label{fig:exp-acc}
\end{figure}

\subsection{Analysis Methods}

We analysed the objective explanation understanding using a Generalised Linear Mixed-Effects Model (GLMM). Each assessment response was coded as correct or incorrect. Accuracy was modelled as a function of \textit{explanation type}, \textit{interaction condition} (Baseline, Scope Disclosure or Explanation Navigator) and \textit{question category} (within-scope vs.\ out-of-scope), together with all interaction terms. We included the participant and assessment item as random intercepts to account for repeated responses from the same participant. We used omnibus Wald tests to examine fixed effects and interaction terms.

To unpack significant effects, we conducted post-hoc Tukey HSD comparisons among the three interaction conditions, separately for each explanation type and question category. We report Tukey-adjusted $p$-values for pairwise comparisons. We additionally conducted pairwise comparisons of overall accuracy across all ten assessment questions.

For subjective ratings of the explanations' ease of understanding and sufficiency of detail, we conducted one-way ANOVAs comparing the three conditions separately for each explanation type. For perceived usefulness of the conversational system, we compared the Scope Disclosure and Explanation Navigator conditions separately for each explanation type using one-way ANOVAs.

For the conversational conditions, we conducted an exploratory analysis of the participants' query activity. Given the volume of interaction data, we used an LLM to code each query by information need and determine whether it was within or beyond the scope of an explanation. As LLM-based coding may introduce classification errors, we treat these results as exploratory. For the Explanation Navigator condition, we additionally used system logs to quantify how often out-of-scope queries triggered routing to complementary explanations.

\subsection{Overall Comprehension}

Table~\ref{tab:glmm_anova} reports the GLMM results. We found a significant main effect of the interaction condition ($\chi^2=255.14$, $p<0.001$), indicating that the level of user comprehension differed across the Baseline, Scope Disclosure and Explanation Navigator conditions. We also found a significant main effect of the question category ($\chi^2=332.43$, $p<0.001$), showing that the participants performed differently on the within-scope and out-of-scope questions. Notably, the interaction condition significantly interacted with the question category ($\chi^2=73.82$, $p<0.001$). This indicates that the benefit of our conversational system differed between the within-scope and out-of-scope questions. We found no significant interactions involving the explanation type, providing no evidence that these patterns differed between local feature importance and ICE.

Pairwise comparisons of the overall accuracy showed that both interventions improved explanation understanding, with the largest gains under the Explanation Navigator. For local feature importance, the Scope Disclosure approach increased accuracy from 49.3\% to 61.5\% ($\Delta=12.2\%$, $p_{\mathrm{adj}}<0.01$), while the Explanation Navigator increased it to 76.0\% ($\Delta=26.7\%$ relative to the Baseline approach, $p_{\mathrm{adj}}<0.001$). The Explanation Navigator also significantly outperformed the Scope Disclosure approach ($\Delta=14.5\%$, $p_{\mathrm{adj}}<0.01$). The same pattern occurred for ICE: the accuracy increased from 40.1\% in the Baseline condition to 49.2\% with the Scope Disclosure approach ($\Delta=9.2\%$, $p_{\mathrm{adj}}<0.05$) and to 72.0\% with the Explanation Navigator ($\Delta=31.9\%$ relative to the Baseline condition, $p_{\mathrm{adj}}<0.001$). The Explanation Navigator again significantly outperformed the Scope Disclosure approach ($\Delta=22.7\%$, $p_{\mathrm{adj}}<0.001$). The full results are reported in Figure~\ref{fig:exp-overall-acc} in Appendix~\ref{app:additional-results}.

In short, these results show that both conversational systems improved the overall comprehension accuracy, with the Explanation Navigator producing the highest performance for both explanation types. However, the significant interaction term between the interaction condition and question category indicates that these improvements arose differently for the within-scope comprehension and the out-of-scope recognition. We therefore examine these two comprehension patterns separately in the following sections.

\subsection{Within-Scope Comprehension}

We first examine the participants' ability to correctly interpret information directly supported by explanations. The results are shown in Figure~\ref{fig:exp-acc}. Across the conditions, the within-scope accuracy was substantially higher than the out-of-scope accuracy, suggesting that the participants were generally able to understand information explicitly given in the explanations.

For local feature importance, Figure~\ref{fig:local-can} shows that both interaction conditions improved the within-scope accuracy as compared to the Baseline approach. The Scope Disclosure technique increased accuracy from 67.6\% to 79.2\% ($\Delta=11.7\%$, $p_{\mathrm{adj}}<0.05$), while the Explanation Navigator increased it further to 80.4\% ($\Delta=12.8\%$, $p_{\mathrm{adj}}<0.05$). However, the Explanation Navigator did not significantly outperform the Scope Disclosure approach. The effect for the ICE explanations is similar. Figure~\ref{fig:ice-can} shows that the Explanation Navigator significantly improved the within-scope accuracy as compared to the Baseline condition ($\Delta=13.4\%$, $p<0.05$), but it did not significantly differ from the Scope Disclosure approach.

These results suggest that both conversational conditions helped the participants interpret the information already given in the explanation. However, the Explanation Navigator did not provide a significant additional benefit over the Scope Disclosure approach for the within-scope comprehension. This is consistent with the system design: when the users' information need was already supported by the current explanation, both conditions operated in the same way, answering the question directly. The distinctive benefit of the Explanation Navigator therefore does not lie in improving the interpretation of within-scope information. %

\subsection{Out-of-Scope Recognition}\label{sec:out-accuracy}

Next, we examine whether the participants recognised when the requested information could not be determined from the displayed explanation. The GLMM shows substantially lower performance on the out-of-scope than the within-scope questions ($\chi^2=332.43$, $p<0.001$). The results in Figures~\ref{fig:local-cannot} and~\ref{fig:ice-cannot} further confirm that in the Baseline condition, the participants correctly recognised only 28.7\% of the out-of-scope questions for local feature importance and 16.4\% for ICE.

This pattern mirrors our interview findings. The participants were comparatively successful at interpreting information explicitly represented by explanations, yet frequently failed to recognise when the requested conclusion could not be supported. Similar findings were reported in prior work~\cite{xuan2025comprehension}, showing that highly comprehensible explanations are still susceptible to misinterpretation. %

The Scope Disclosure approach substantially improved the out-of-scope recognition. For local feature importance, it increased the out-of-scope accuracy by 21.0 percentage points over the Baseline condition ($p_{\mathrm{adj}}<0.01$); for ICE, it increased accuracy by 17.9 percentage points ($p_{\mathrm{adj}}<0.01$). These improvements confirmed our DC2 from Section~\ref{sec:design-considerations}: making a relevant scope explicit can help the users recognise that the information they seek is unsupported.

However, the Scope Disclosure approach alone did not fully resolve the problem. The participants still failed to recognise approximately half of the out-of-scope questions for local feature importance and almost two-thirds for ICE. Making the scope explicit therefore appears useful but insufficient: it explains why the current inference is unsupported, while leaving the users' original information needs unresolved.

Our Explanation Navigator produced an additional improvement beyond that offered by the Scope Disclosure technique. For local feature importance, the Explanation Navigator achieved the out-of-scope accuracy of 65.7\%, significantly exceeding both the Baseline ($\Delta=37.0\%$, $p_{\mathrm{adj}}<0.001$) and Scope Disclosure ($\Delta=16.0\%$, $p_{\mathrm{adj}}<0.05$) approaches. 
The incremental effect was even larger for ICE: the Explanation Navigator increased the out-of-scope accuracy to 60.6\%, a 44.2-point improvement over the Baseline condition ($p_{\mathrm{adj}}<0.001$) and a 26.3-point increase over the Scope Disclosure approach ($p_{\mathrm{adj}}<0.001$).

These results show the distinctive benefit of our Explanation Navigator. The Scope Disclosure technique helps users to recognise why the current explanation cannot support an inference; the Explanation Navigator additionally directs them to an explanation that can provide the information they seek. The significant improvement over the Scope Disclosure approach suggests that combining the explicit scope clarification with complementary explanations better supports the users in distinguishing what can and cannot be inferred from the original explanation.

\begin{figure}
    \centering
    \begin{subfigure}{0.9\linewidth}
        \centering
        \includegraphics[width=\linewidth]{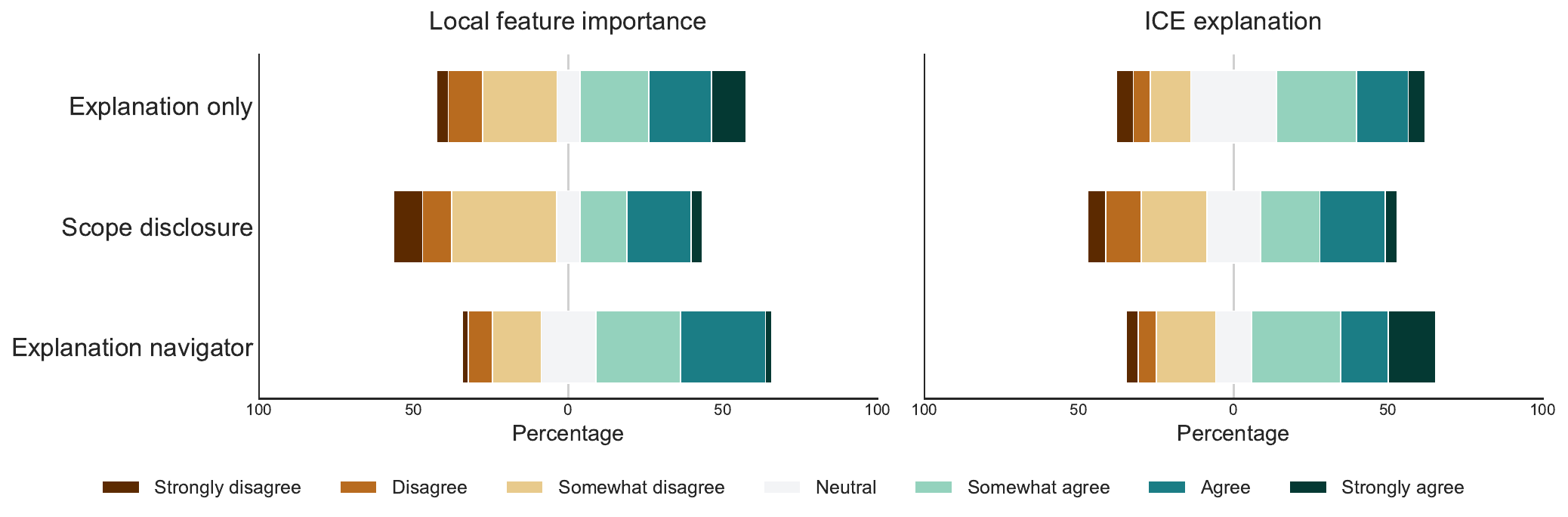}
        \caption{Likert scale responses to the ``this explanation is sufficiently detailed'' question stratified by the interaction condition for the local feature importance (left) and ICE (right) explanations.}
        \label{fig:sub-detailed}
    \end{subfigure}
    \\
    \begin{subfigure}{0.8\linewidth}
        \centering
        \includegraphics[width=\linewidth]{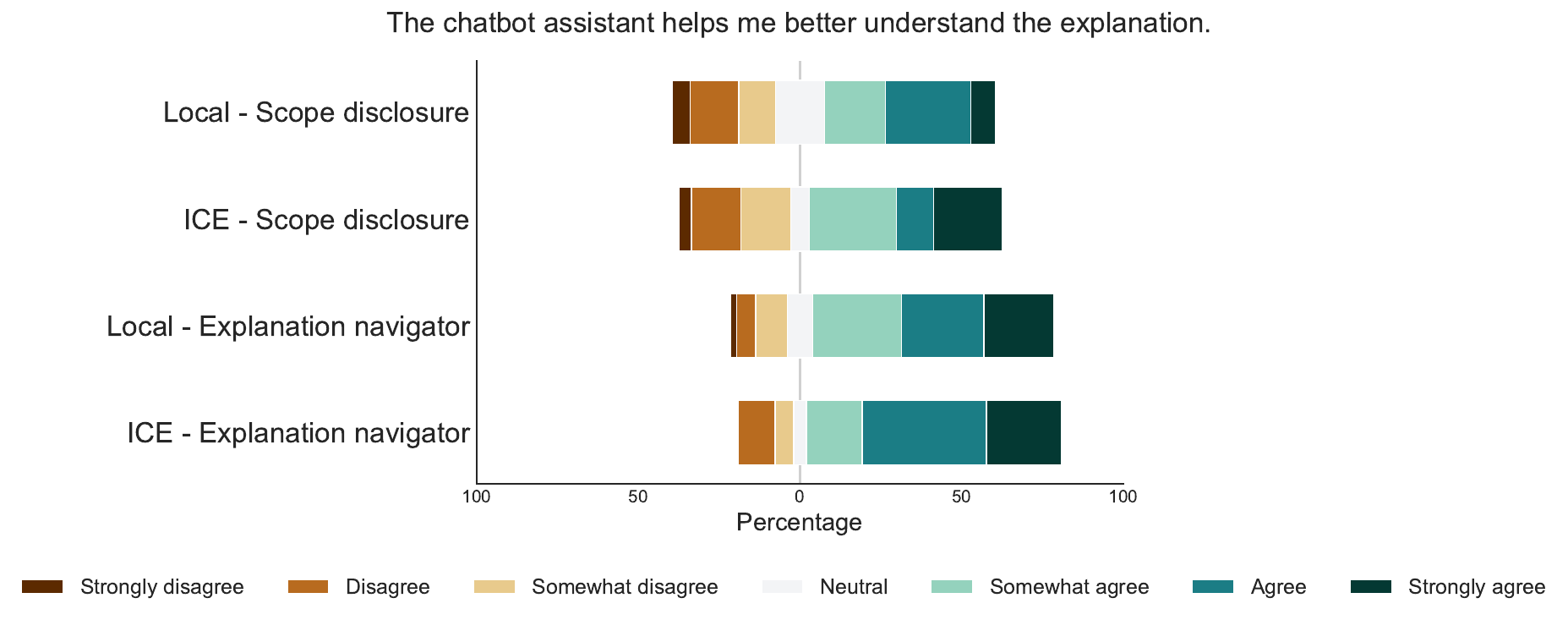}
        \caption{Likert scale responses to the ``the chatbot assistant helps me better understand the explanation'' question.}
        \label{fig:sub-chatbot}
    \end{subfigure}
    \caption{Likert score response to our subjective assessment questions.}
    \label{fig:subjective-rating}
\end{figure}

\subsection{Subjective Assessment}

We additionally examine whether the interaction conditions changed the participants' subjective perception of the explanations. For both explanation types, the Explanation Navigator received numerically higher ratings for the \emph{perceived ease of understanding}, but the differences are not statistically significant. These results are reported in Figure~\ref{fig:sub-easy-understand} in Appendix~\ref{app:additional-results}. 

A different pattern emerges for the \emph{perceived sufficiency of detail}. For local feature importance, the participants rated the explanation as significantly less detailed under the Scope Disclosure approach than under the Baseline condition ($p<0.05$), while the Explanation Navigator was rated significantly higher than the Scope Disclosure technique ($p<0.05$). No significant differences can be observed for ICE. We hypothesise that the Scope Disclosure approach made the participants more aware of the information that the current explanation did not provide, whereas the Explanation Navigator supplemented this clarification with additional explanatory information.

The participants in the two conversational conditions also rated the \emph{usefulness of the tool} for helping them better understand the explanation. The Explanation Navigator was rated significantly more useful than the Scope Disclosure approach for both local feature importance ($p=0.01$) and ICE ($p<0.05$). This suggests that the participants valued not only being told what the current explanation could not support but also being directed to explanatory information that addresses their question.

Overall, the subjective results complement our objective findings. The Explanation Navigator did not significantly change the perceived ease of understanding, but the participants consistently found it more useful than the Scope Disclosure approach. For local feature importance, it also restored the perceived sufficiency of detail relative to the disclosure alone. Therefore, exposing an explanation scope may make its incompleteness more salient, while coupling that clarification with complementary information can preserve the explanations' perceived usefulness.

\subsection{Participant Interaction Patterns}

We further analysed the participants' interaction logs to understand how they used the conversational systems. 
The participants in the Scope Disclosure and Explanation Navigator conditions spent an average of 6.42 and 7.39 minutes interacting with the systems and submitted 11.4 and 12.1 queries respectively.

Across both conditions, 77.5\% of the queries were out-of-scope for the displayed explanation, indicating that the participants frequently sought information the current explanation could not support. For local feature importance, the most common out-of-scope need concerned \emph{action} (14.1\%), whereas for ICE it concerned \emph{feature ranking} (19.2\%). %

In the Explanation Navigator condition, 65.2\% of the out-of-scope queries triggered a complementary explanation; the remainder primarily concerned terminology clarification or broader questions about the AI model. 
These interaction patterns complement the accuracy results reported in Section~\ref{sec:out-accuracy}: scope mismatches often emerged during the participants' own exploration and the Explanation Navigator frequently redirected these information needs to explanations with appropriate scope. This provides further evidence for its benefit in helping the users resolve information needs that extend beyond the current explanation.

\section{Discussion}\label{sec:discussion}

This work unpacks the reasoning process governing explainees' out-of-scope interpretations of AI explanations and introduces the Explanation Navigator as a strategy for overcoming it. In this section, we discuss how the notion of leaky explanations extends existing accounts of explanation misunderstanding, what the Explanation Navigator contributes to the design of XAI systems and also how our findings reframe the evaluation of explanation understanding.

\subsection{Leaky Explanations Establish a Mechanism Underlying Explanation Misunderstanding}%

Prior human-centred XAI research has documented several distinct forms of mismatch between what explanations technically represent and what explainees understand about them. \citet{chromik2021think} found that users generalised local feature importance to global model behaviour. \citet{collaris2022characterizing} showed that data scientists' expectations of local feature importance can conflict with method assumptions. \citet{xuan2025comprehension} found that users could understand explicitly presented information while still inferring unspecified information. 
These studies establish that explanation misunderstanding is widespread. Our findings further identify a recurring reasoning process through which such misunderstanding can arise.

We characterise explainees' reasoning process through the notion of \emph{leaky explanations}. An explanation becomes leaky when the properties hidden by its employed abstraction become necessary for determining what can legitimately be inferred from it. 
When explainees seek information that falls within the explanation's scope, knowledge of these hidden properties may be unnecessary; thus our participants could often correctly interpret what the explanations represented explicitly. 
However, when explainees seek information beyond that scope, determining whether the desired inference was valid requires knowledge of the hidden properties that give rise to the scope. 
Without access to these properties, our participants often substituted their own assumptions and used them to derive an answer, rather than recognising that the requested information is unavailable. Misunderstanding therefore arose not from the explainees' inability to read the explanation, but from introducing assumptions that diverged from the explanations' hidden properties necessary for correct interpretation.

Our conceptualisation of \emph{leakiness} differs from several related accounts of how misunderstanding arises. 
First, an explanation may have a \emph{limited scope}, which restricts what information it can represent~\cite{dwivedi2023explainable}; leakiness, on the other hand, concerns what happens when the users' information needs cross that boundary, with correct interpretation depending on properties hidden from the explainees' view. The limited scope alone therefore does not make an explanation leaky. The same explanation can be sufficient for one information need yet become leaky for another. 
Second, the explainees' \emph{inaccurate mental models} capture the resulting state of their understanding~\cite{kulesza2013too}, whereas leakiness describes a process through which that state may be constructed. Finally, leakiness is different from the explainees' \emph{cognitive biases}~\cite{wang2019designing}; cognitive biases can influence which assumptions the users introduce, but leakiness pertains to when such assumptions become consequential because relevant explanatory properties are hidden by the explanation itself.

Leakiness therefore does not solely attribute misunderstanding to a deficient mental model or an intrinsic weakness of an explanation. It emerges from the interplay of \emph{what the explanation represents}, \emph{what information the users seek} and \emph{what assumptions the users introduce to connect the two}. Our perspective shifts attention from making individual explanations universally more complete and towards helping explainees recognise when their reasoning has moved beyond what a particular explanatory artefact can support.

\subsection{Scope Disclosure and Complementary Explanations Help Rectify Misunderstanding}

The mechanism of leaky explanations also clarifies why simply supplying more explanatory information may be insufficient. Once the users have constructed an unsupported interpretation, providing the information they were seeking does not necessarily demonstrate why their original inference was invalid. This pattern appeared in both our dashboard and initial conversational explainer stages, where the participants could access an appropriate explanation while retaining their earlier interpretation. 
The Scope Disclosure approach directly addresses the out-of-scope interpretations by making their scope explicit. %
However, the disclosure alone remained insufficient as the out-of-scope accuracy under this condition was only 49.7\% for local feature importance and 34.3\% for ICE.

We argue that the remaining out-of-scope errors arise because the scope disclosure addresses only one part of the mismatch between the explanations and the users' information needs. It tells the users why the current explanation cannot answer their question but does not indicate where the answer can be found. For example, telling a user that the slope of an ICE curve cannot establish feature importance left the question of feature importance unresolved.

Our Explanation Navigator solves this unmet need by complementing the scope disclosure with a complementary explanation. Consequently, it produces a further significant improvement in out-of-scope recognition for both explanations. We suggest that the complementary explanation establishes a \emph{contrast} to the scope of the original explanation. This contrast helps the users revise not only their answers to the immediate questions but also their understanding of the scope of the original explanation. 

The scope disclosure and complementary explanations are therefore reinforcing corrective mechanisms rather than alternative strategies. Scope disclosure addresses the unsupported inference, while the complementary explanations resolve the information needs that gave rise to it. This is consistent with knowledge-revision and refutation research, which shows that misconceptions are more effectively revised when an existing conception is explicitly challenged and a coherent alternative is provided~\cite{kendeou2014krec,kendeou2019knowledge,schroeder2022refutation,rich2017belief}.

\subsection{Explanation Navigator Rectifies Misinterpretations, Not Just Selects Complementary Explanations}

Interactive and conversational XAI increasingly treats explainability as an iterative process in which the users can request additional explanations for their evolving information needs~\cite{bertrand2023selective,slack2023explaining}. %
The framework of our Explanation Navigator addresses a different interaction problem that arises once a user has already begun interpreting an explanation. At this point, a follow-up question is not just a request for new information; it may also reveal an unsupported inference about the explanation currently being viewed. If a user asks ``Which feature is most important?'' while viewing an ICE explanation, 
the question may indicate that they believe the properties of the ICE curve, such as its slope, can be used to infer feature importance.  %

The Explanation Navigator therefore treats the current explanation as part of the conversational context. Rather than mapping only from a question to the next explanation, it reasons over the relationship between the users' information needs and the explanations from which these needs emerge. When the two are misaligned, it first addresses the misinterpretation of the current explanation and only then proceeds to complementary information. 
This distinction extends the conversational XAI paradigm from \emph{explanation selection} toward \emph{interpretation rectification}. It fulfils earlier calls for interactive XAI to account for the explainees' evolving understanding~\cite{chromik2021think,rutjes2019considerations}, while identifying their subsequent information needs as a signal through which the potential scope of misunderstandings can be detected and addressed.

\subsection{Implications for Human-Centred XAI Evaluation}

Leaky explanations also expose a gap in how explanation understanding is commonly evaluated. The existing human-centred XAI measures assess dimensions such as perceived explanation quality, model understanding and downstream task performance~\cite{hoffman2023measures,rong2023towards,kim2024human}. Comprehension measures often test whether explainees can correctly interpret information explicitly represented by an explanation~\cite{cheng2019explaining,wang2021explanations}. Such measures can establish that the users understand what is shown but may fail to detect whether they also derive conclusions that the explanation does not support.

The results of our quantitative study suggest that these two aspects of understanding should be treated distinctly. Our participants performed substantially better on within-scope than out-of-scope questions, indicating that correctly interpreting displayed information does not imply recognising the limits of what can be inferred from it -- a phenomenon observed earlier in the XAI literature~\cite{xuan2025comprehension}. Evaluating only the within-scope understanding therefore risks treating an explanation as successfully understood even when its users make unsupported inferences beyond its scope.

We therefore recommend incorporating deliberately out-of-scope questions into human-centred XAI evaluation that are accompanied by a valid ``cannot tell from this explanation'' response as proposed earlier by \citet{xuan2025comprehension}. Such questions test whether explainees understand the boundaries of correct insights rather than only the visible content of the explanations. This measure should complement, rather than replace, existing approaches to task performance and model understanding evaluation~\cite{poursabzi2021manipulating,bhattacharya2024exmos}. It provides an intermediate check on whether successful task performance is grounded in a valid interpretation of the explanation or achieved despite actually misunderstanding its scope.

\subsection{Limitations and Future Work}

Our work has several limitations. First, our quantitative study evaluated the Explanation Navigator with two local explanations. Although our qualitative findings suggest similar challenges for counterfactual explanations, future work should test whether our approach generalises to other local and global explanation types. Second, our evaluation focused on addressing misunderstanding pertaining to a single explanation. Since our interview study showed that providing multiple explanations does not automatically clarify their distinct scopes, future work could examine how the Explanation Navigator can help explainees to identify which explanation is appropriate for a given information need. 
Lastly, our evaluation used a credit scoring scenario with primarily non-technical participants. Prior work has shown that expert users can also form inaccurate assumptions about explainability methods~\cite{collaris2022characterizing,kaur2020interpreting} but the relevant explanation scope may differ by expertise and domain. Future studies should therefore evaluate our Explanation Navigator with expert users and in other high-stakes settings.

\section{Conclusion}

This work examined how explanation misunderstanding emerges when explainees seek information beyond what an AI explanation can offer. We found that the users deal with such unavailable explanation scope by confabulating assumptions about the omitted aspects of explanations and using them to arrive at unsubstantiated insights -- a mechanism we termed \emph{leaky explanations}. We further showed that simply providing additional explanations does not necessarily rectify these incorrect interpretations. To address this problem, we introduced the \emph{Explanation Navigator}, which makes the relevant explanation scope explicit before directing explainees to complementary explanatory information. Our quantitative study demonstrated that this approach effectively rectified the users' out-of-scope misunderstanding. More broadly, this work highlights the importance of designing XAI systems that help explainees understand not only what an explanation communicates but also where its explanatory scope ends.

\section*{Acknowledgement}
This research was conducted by the ARC Centre of Excellence for Automated Decision-Making and Society (Project No.\ CE200100005), funded by the Australian Government through the Australian Research Council. This work was partially funded by an unrestricted gift from Google under the Research Scholar Program. The computing resources were provided by the RMIT Advanced Computing Ecosystem. Additional support was provided by the TRUST-ME project (205121L 214991), funded by the Swiss National Science Foundation. %

\bibliographystyle{ACM-Reference-Format}
\bibliography{references}

\appendix
\section{User Study Materials and Interfaces}\label{app:survey}

This appendix provides the materials and interfaces presented to the participants in the interview and online study.

\begin{figure}[h]
    \centering
    \includegraphics[width=0.5\linewidth]{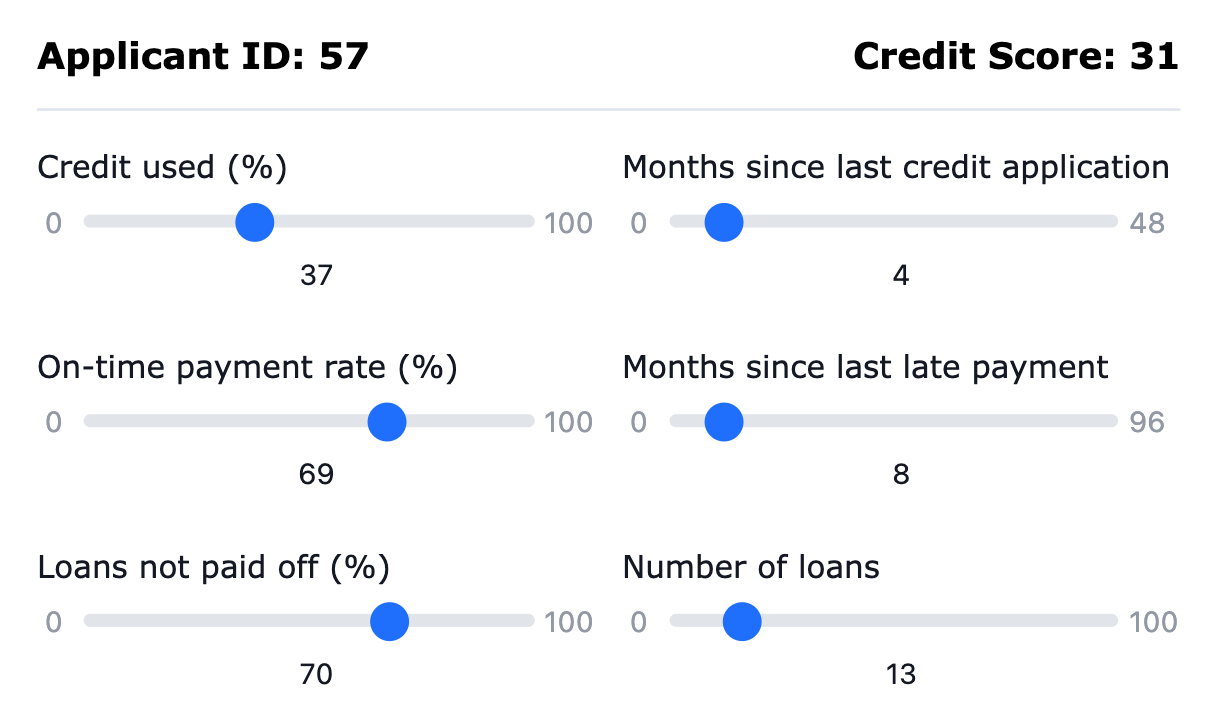}
    \caption{Applicant profile and AI prediction shown in the interview study. For ease of reference, this applicant is referred to as Alex. The meaning of each feature was explained verbally to the participants during the interview.}
    \label{fig:alex-profile}
\end{figure}

\begin{figure}[h]
    \centering
    \includegraphics[width=0.8\linewidth]{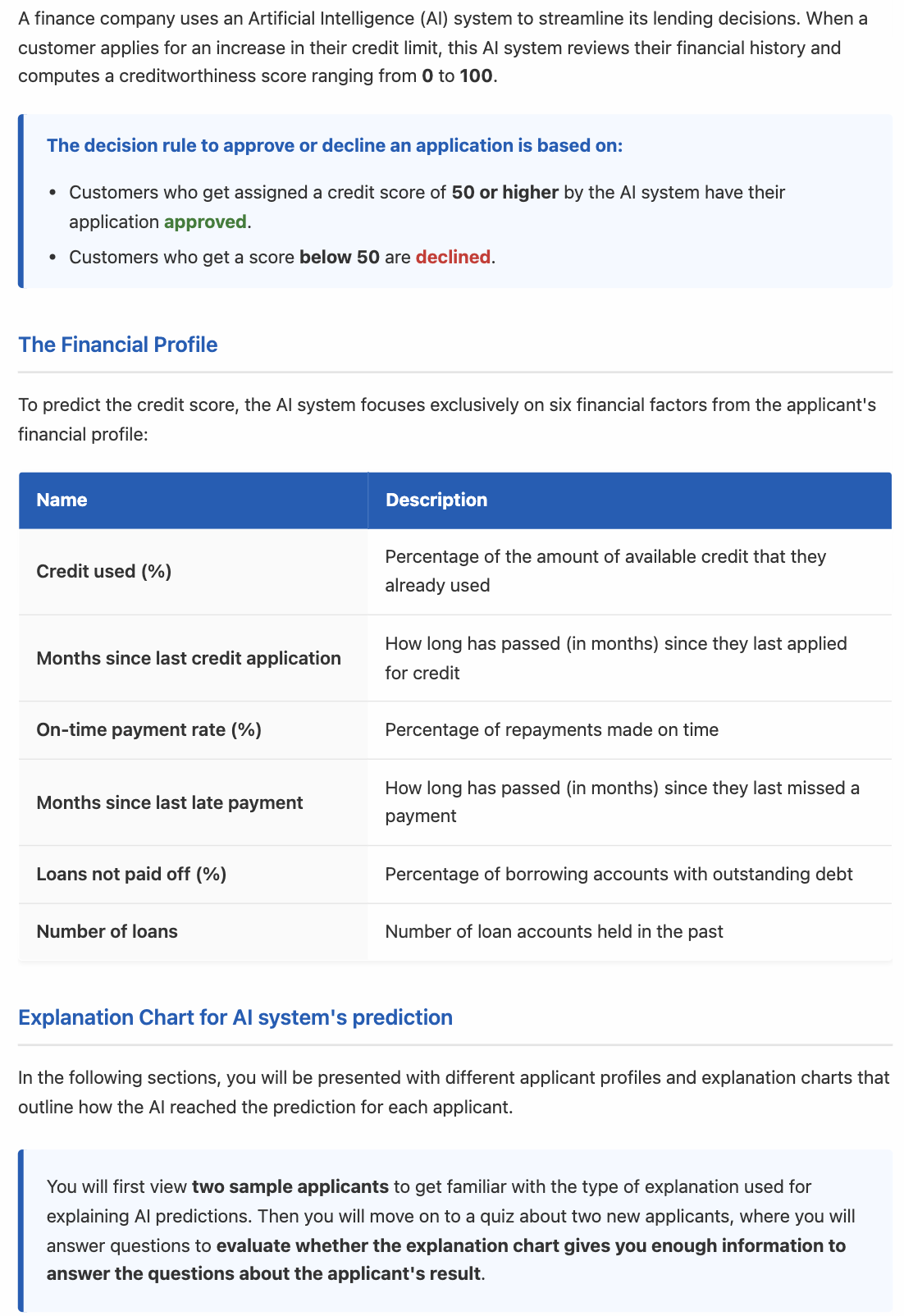}
    \caption{Introduction page of the online survey study describing our credit limit increase scenario, the role of the AI system and task instructions. This page was shown to the participants in all experimental conditions.}
    \label{fig:survey-intro}
\end{figure}

\begin{figure}[h]
    \centering
    \includegraphics[width=0.8\linewidth]{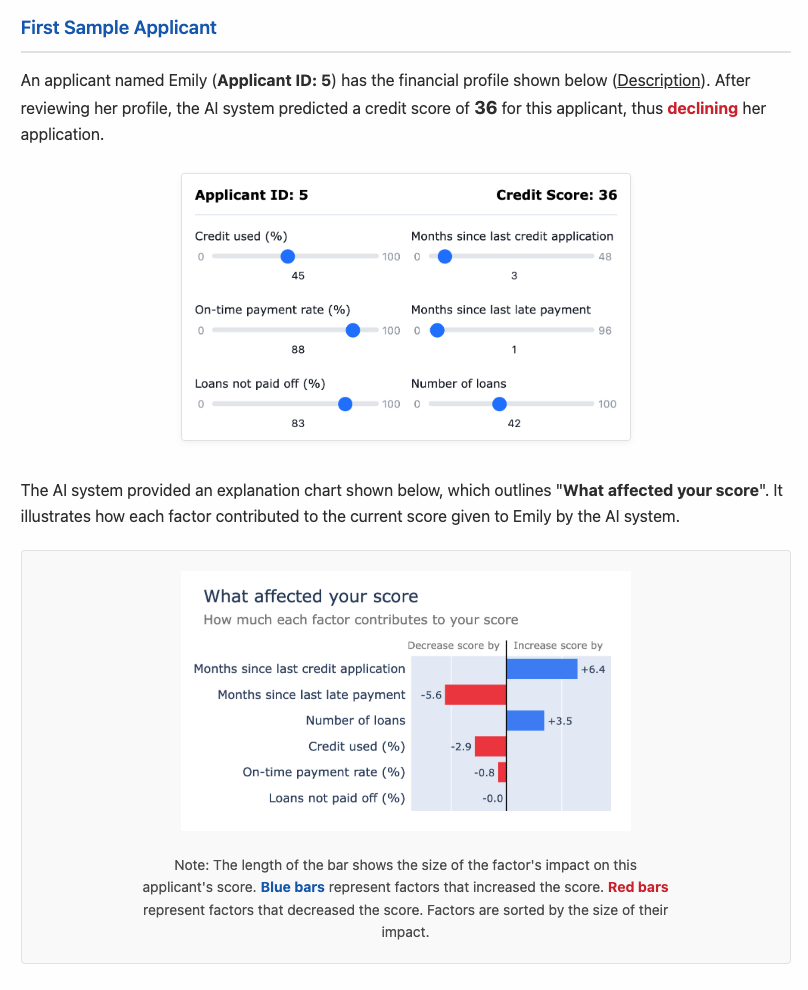}
    \caption{Example local feature importance explanation for a sample applicant shown during the \emph{exposure phase}. This sample applicant is referred to as Emily. The sample applicants used for exposure were distinct from the test applicants used in the main assessment. This page was shown to the participants in all experimental conditions.}
    \label{fig:sample-applicant-intro}
\end{figure}

\begin{figure}[h]
    \centering
    \includegraphics[width=0.8\linewidth]{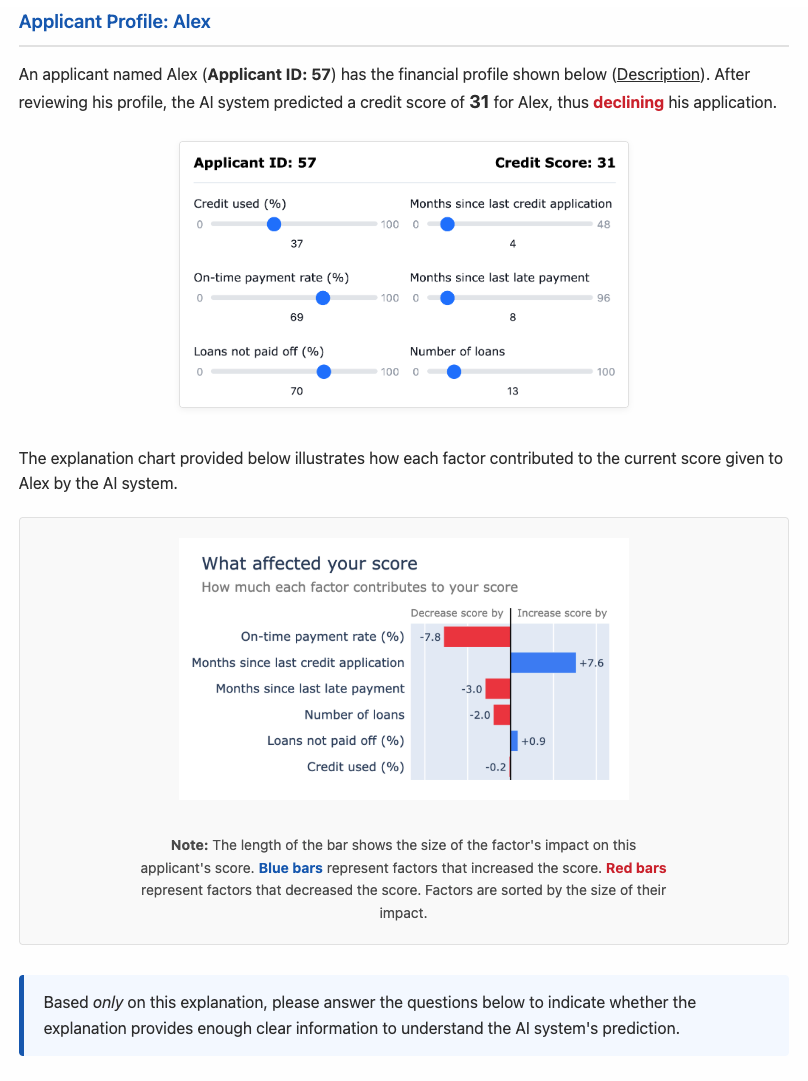}
    \caption{Local feature importance explanation shown for one test applicant during the \emph{main assessment} phase. This test applicant is referred to as Alex. The test applicants were distinct from the sample applicants used during the exposure phase.}
    \label{fig:test-applicant-intro}
\end{figure}

\begin{figure}[p]
    \centering
    \includegraphics[height=0.9\textheight, width=\linewidth, keepaspectratio]{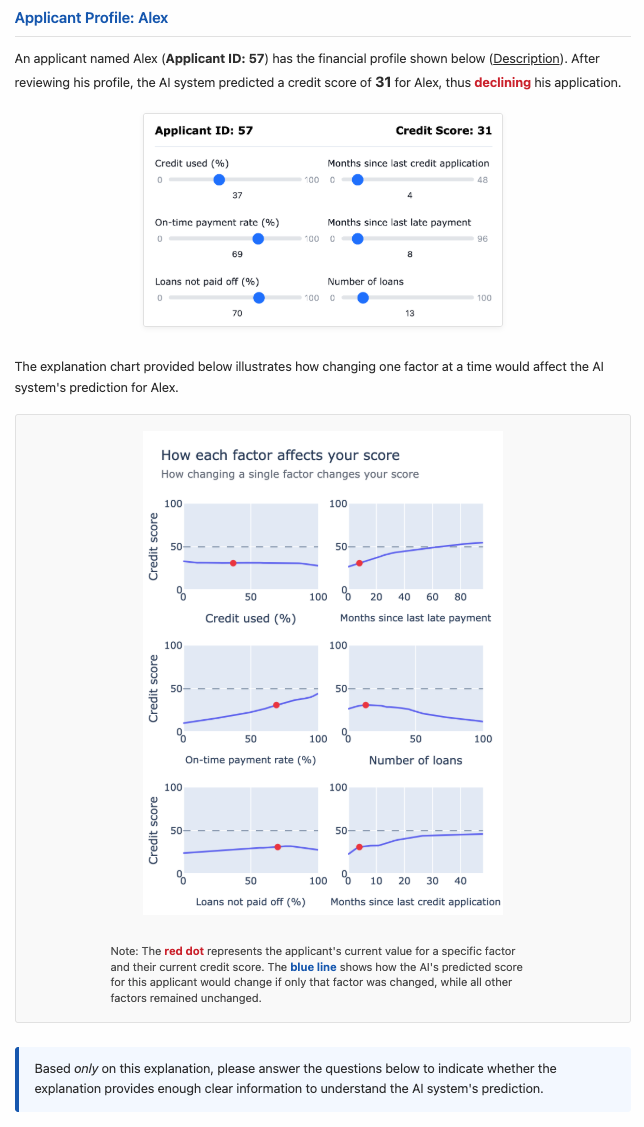}
    \caption{ICE explanation shown for one test applicant during the main assessment phase. This test applicant is referred to as Alex. The interface was otherwise identical to the local feature importance condition shown in Figure~\ref{fig:test-applicant-intro}.}
    \label{fig:test-applicant-ice}
\end{figure}

\begin{figure}[h]
    \centering
    \begin{subfigure}[b]{0.48\linewidth}
        \includegraphics[width=\linewidth]{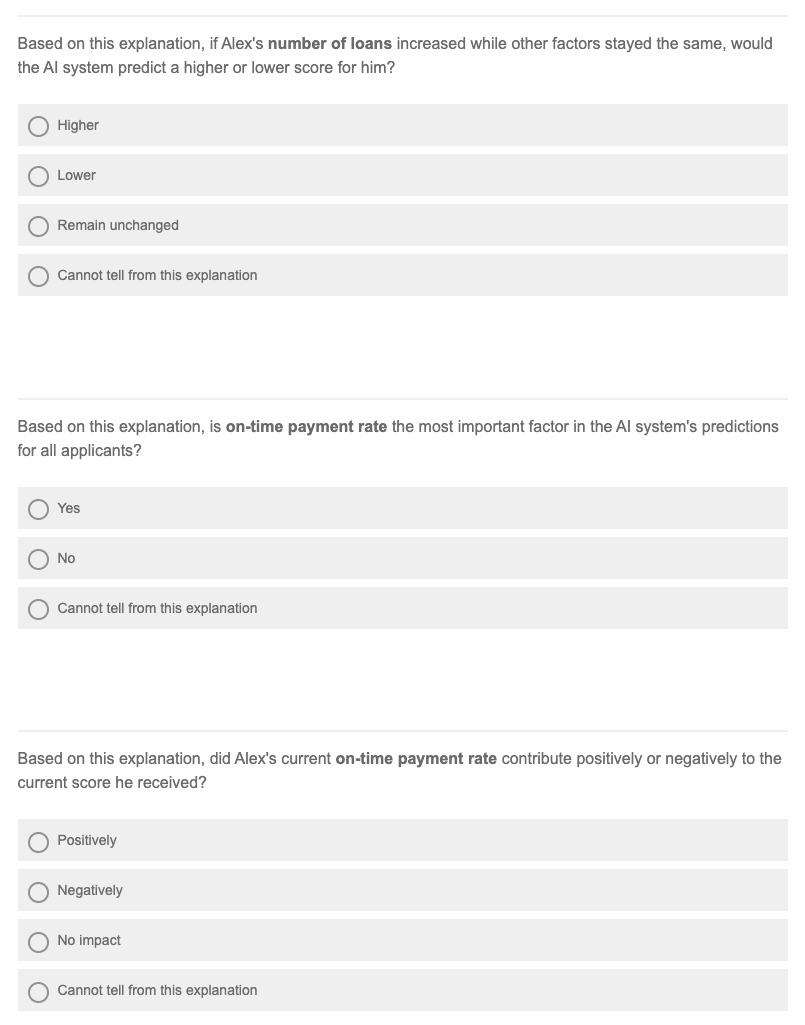}
        \caption{Feature direction, model-level, feature effect questions.}
        \label{fig:q_list1}
    \end{subfigure}
    \hfill
    \begin{subfigure}[b]{0.48\linewidth}
        \includegraphics[width=\linewidth]{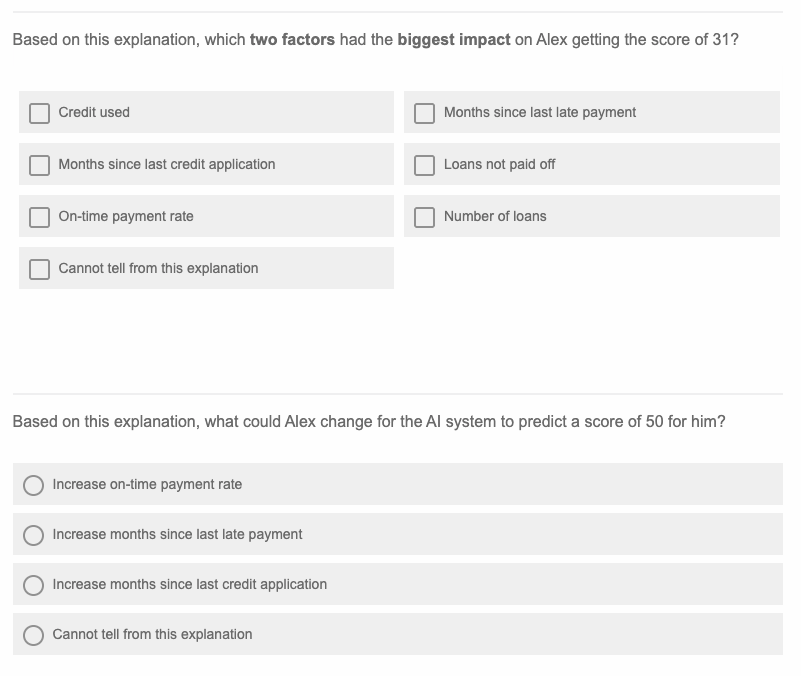}
        \caption{Feature ranking and action questions.}
        \label{fig:q_list2}
    \end{subfigure}
    \caption{Five assessment questions presented for a test applicant (Alex) in the \emph{local feature importance} condition. The question order was randomised for each participant.}
    \label{fig:objective-questions}
\end{figure}

\begin{figure}[h]
    \centering
    \begin{subfigure}[b]{0.48\linewidth}
        \includegraphics[width=\linewidth]{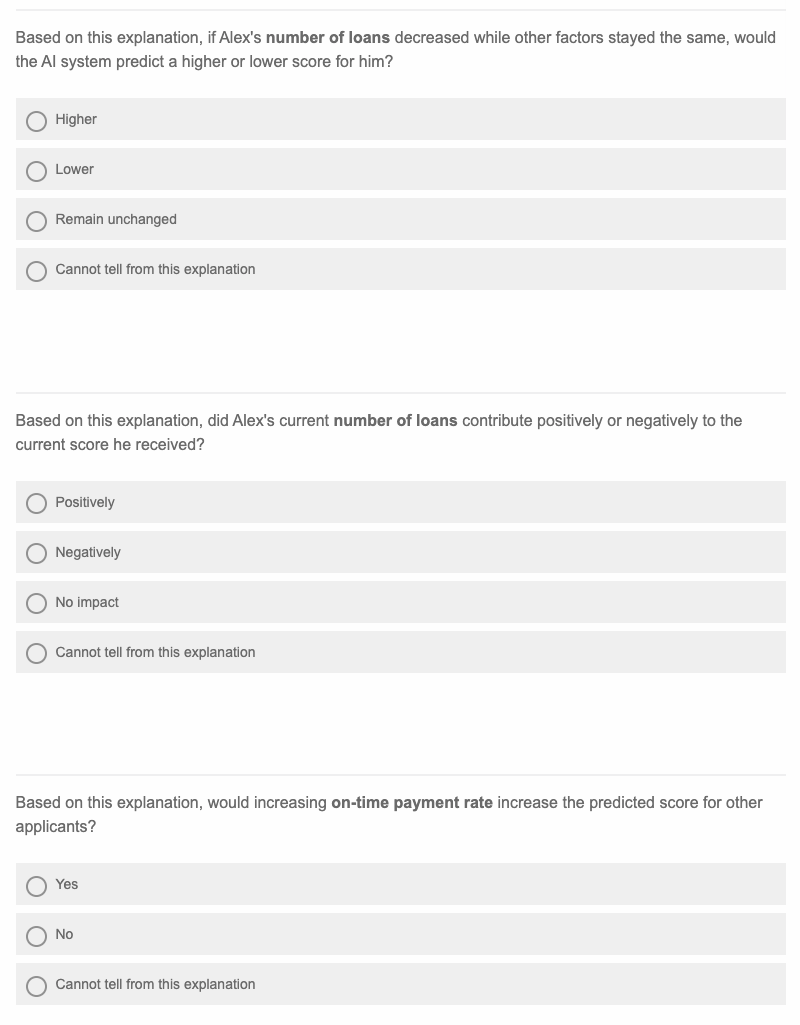}
        \caption{Feature direction, feature effect, and model-level questions.}
        \label{fig:q_list1_ice}
    \end{subfigure}
    \hfill
    \begin{subfigure}[b]{0.48\linewidth}
        \includegraphics[width=\linewidth]{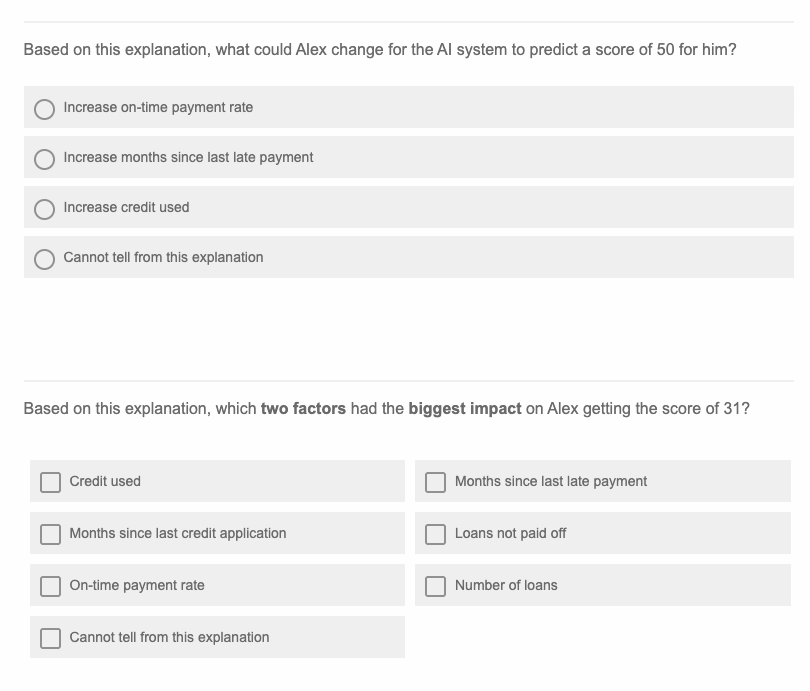}
        \caption{Action and feature ranking questions.}
        \label{fig:q_list2_ice}
    \end{subfigure}
    \caption{Five assessment questions presented for a test applicant (Alex) in the \emph{ICE} condition. The question order was randomised for each participant.}
    \label{fig:objective-questions-ice}
\end{figure}

\begin{figure}[p]
    \centering
    \includegraphics[height=0.9\textheight, width=\linewidth, keepaspectratio]{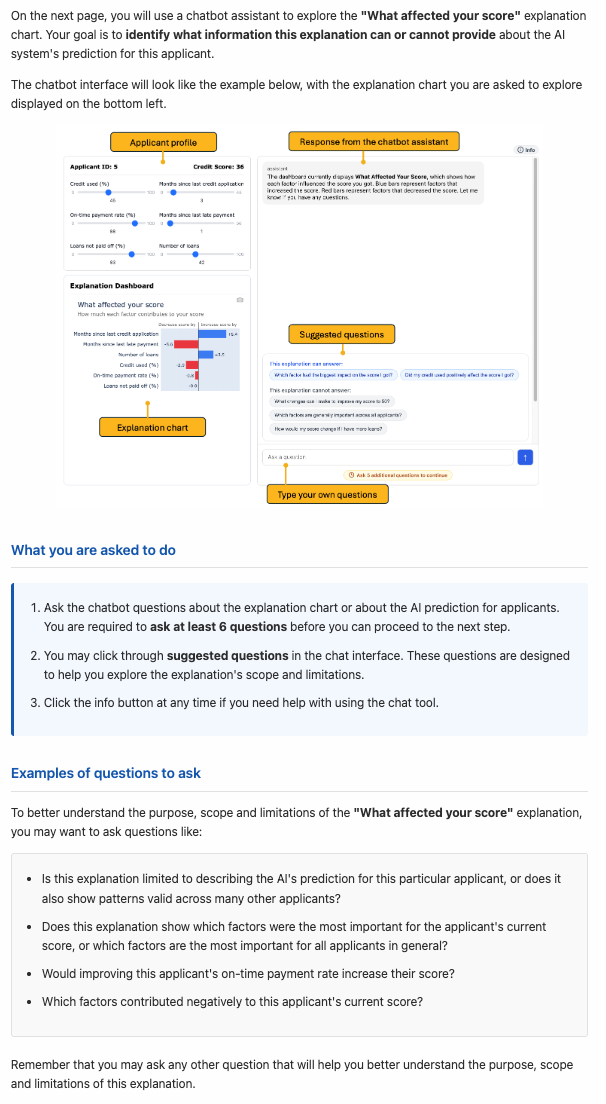}
    \caption{Introduction to the conversational system. This page was shown only to the participants in the Scope Disclosure and Explanation Navigator conditions.}
    \label{fig:llm-interface}
\end{figure}

\begin{figure}[h]
    \centering
    \includegraphics[width=\linewidth]{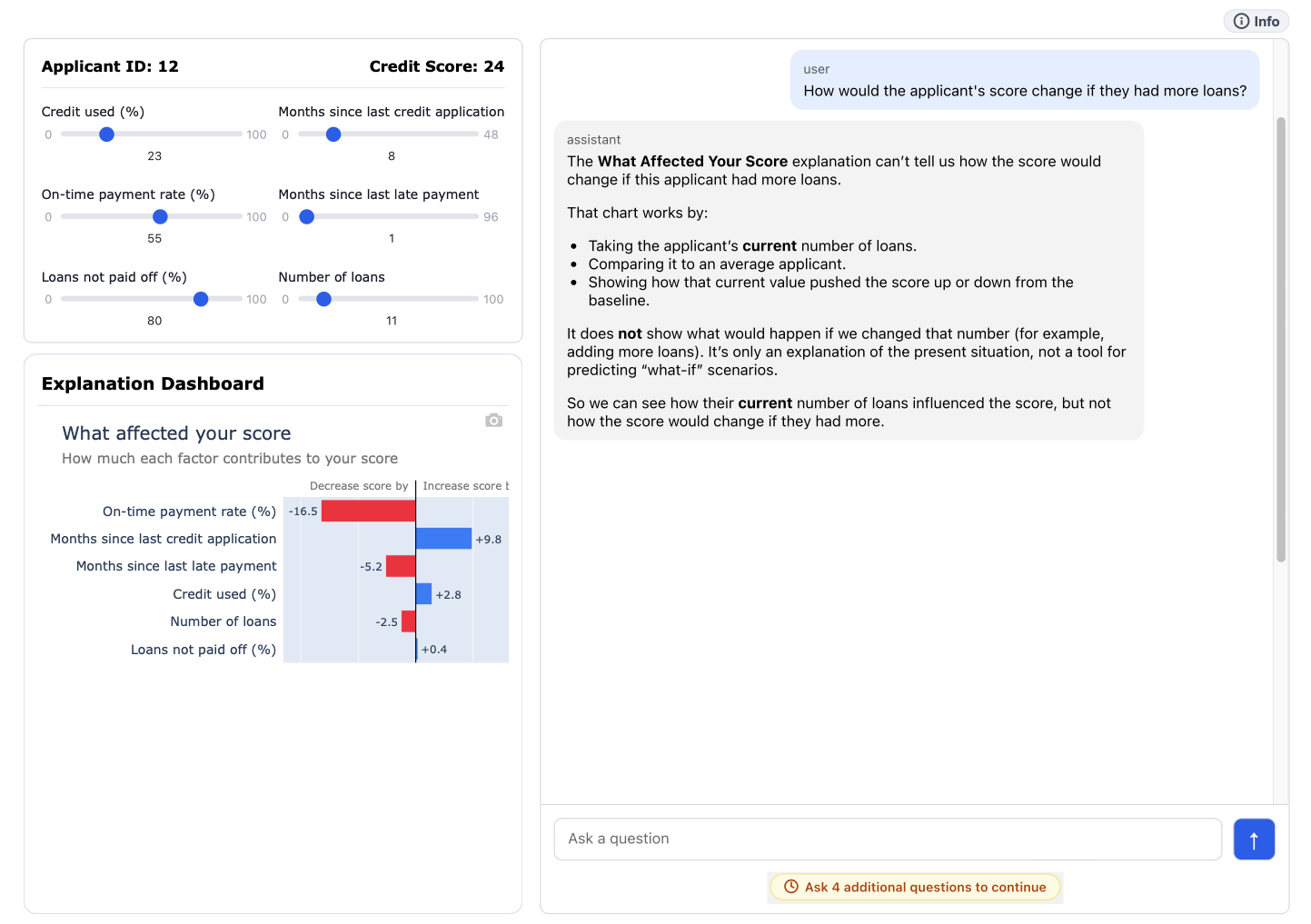}
    \caption{Example response produced in the Scope Disclosure condition. When the user's information need exceeded the scope of the current explanation, the system made the relevant scope explicit and explained why the requested inference was unsupported without providing a complementary explanation.}
    \label{fig:scope-disclosure-response}
\end{figure}

\clearpage\newpage

\section{Technical Details of the Explanation Navigator}\label{app:tech-detail}

\begin{table}[h]
    \caption{Current functions and XAI tools supported in our execution layer.}
    \label{tab:tool-desc}
    \centering
    \small
    \begin{tabular}{@{}p{0.47\linewidth}p{0.5\linewidth}@{}}
    \toprule
        Function definition & Description \\ \midrule
        \verb|generate_local_shap_bar_plot(instance_id)|  & Compute local feature importance for a specific instance \\
        \verb|generate_global_subgroup_shap_plot(source, indices)| & Compute global or subgroup feature importance \\
        \verb|get_individual_prediction(instance_id)| & Retrieve model prediction for a specific instance \\
        \verb|get_average_prediction(source, indices)| & Compute average prediction for a dataset or subgroup \\
        \verb|get_ice_plot(instance_id, feature)| & Generate ICE plot for a specific feature \\
        \verb|get_counterfactual_explanation(instance_id)| & Compute smallest feature changes leading to a different prediction \\
        \verb|get_subgroup(filters)| & Filter out instances according to feature or prediction criteria \\
        \verb|predict_with_feature_changes(instance_id, changes)| & Recompute prediction after applying feature changes \\
        \verb|get_similar_instances(instance_id, k)| & Retrieve $k$ most similar instances \\
        \verb|get_representative_instances(indices, k)| & Retrieve $k$ representative instances from a given group \\
        \verb|dataset_meta()| & Retrieve metadata describing the dataset and its features \\
        \verb|model_meta()| & Retrieve metadata describing the AI model and its output \\
    \bottomrule
    \end{tabular}
\end{table}

\subsection{Implementation Details}

Our conversational Explanation Navigator was implemented using GPT-5.1 developed by OpenAI~\cite{openai2025gpt}. We set the temperature to 1.0 for the \emph{detection} layer and 0.1 for the \emph{response} layer. The system was deployed on Microsoft Azure App Service. The explanation tools in the \emph{execution} layer were implemented in Python; the frontend interface was developed in JavaScript. The set of tools supported by the \emph{execution} layer is summarised in Table~\ref{tab:tool-desc}. A demonstration of our deployed prototype is provided in the supplementary video material.

\subsection{Technical Evaluation}

Across 103 participants assigned to the Explanation Navigator condition, we collected 1251 natural language queries and the corresponding tool calls produced by the conversational system. To assess whether the \emph{detection} layer correctly mapped the users' information needs to the appropriate explanation operation, we randomly sampled 100 queries for manual evaluation. 
For each sampled query, we examined whether the LLM selected the tool that was appropriate for addressing the users' questions. The resulting tool-selection accuracy was 83\%. 

\subsection{Explanation Knowledge Base}\label{app:knowledge-base}

\begin{lstlisting}[
    style=promptstyle,
    caption={Explanation \emph{knowledge base} for two explanation methods: local feature importance and ICE.},
    label={lst:function_desc},
    captionpos=t      %
]
{
"type": "function",
"name": "generate_local_shap_bar_plot",
"description": "Generates 'What affected your score' chart for ONE specific applicant. Shows what factors pushed ONE specific applicant's predicted score up or down and by how much. 
    Use this when the user asks things like 'why is my score high/low?', 'what factor was the most important for my application?', 'did my number of loans positively/negatively affect my score?'.
    CRITICAL: If the user asks 'Why is factor X the most important?' or 'What is the most important factor?' while discussing their own profile or score, they are asking about LOCAL importance. Use this tool.
    This explanation applies only to the selected applicant and does NOT represent factor importance across applicants.
    Generation: This chart is generated by comparing your current details to an average applicant and see how much each factor moved the system's prediction from its baseline to your score. Because it looks at your current profile, it explains your current score but does not predict what happens if you change your factors."
},
{
"type": "function",
"name": "get_ice_plot",
"description": "Generates 'How changing one factor affects your score' chart for a specific applicant. Shows how that applicant's predicted credit score changes when changing a single factor while all other factors are kept fixed at their original values. 
    Use this tool when the user asks:
        - 'How changing this factor changes my score?'
        - 'What happens if I lower my Number of loans'
        - 'What if this factor was higher or lower?'
    Note: This explanation only applies for a specific applicant. Feature must match a valid factor name exactly.
    Generation: This chart is generated by keeping all your details exactly the same while testing how the predicted score would change when only changing one factor. Because it only tests one feature at a time not how much a feature matters relative to other features, it cannot rank which features were most important for your current score."
}
\end{lstlisting}

\clearpage\newpage

\section{Additional Results for Quantitative Study}\label{app:additional-results}

\subsection{Participants Demographic Information}

\begin{table}[h]
\caption{Participant demographics and domain familiarity ($N = 316$).}
\label{tab:demographics}
\centering
\small
\begin{tabular}{lrr}
\toprule
\textbf{Category / Response} & \textbf{Count ($n$)} & \textbf{Percentage (\%)} \\
\midrule
\multicolumn{3}{l}{\textbf{Age}} \\
\quad 18--29 years old & 85 & 26.9 \\
\quad 30--44 years old & 145 & 45.9 \\
\quad 45--59 years old & 61 & 19.3 \\
\quad 60+ years old & 25 & 7.9 \\
\midrule
\multicolumn{3}{l}{\textbf{Gender}} \\
\quad Female & 157 & 49.7 \\
\quad Male & 154 & 48.7 \\
\quad Prefer not to say & 4 & 1.3 \\
\quad Non-binary / third gender & 1 & 0.3 \\
\midrule
\multicolumn{3}{l}{\textbf{AI Familiarity}} \\
\quad No knowledge & 2 & 0.6 \\
\quad Negligible knowledge & 58 & 18.4 \\
\quad Some knowledge & 165 & 52.2 \\
\quad Moderate knowledge & 80 & 25.3 \\
\quad Extensive knowledge & 11 & 3.5 \\
\midrule
\multicolumn{3}{l}{\textbf{Credit Application Familiarity}} \\
\quad No knowledge & 26 & 8.2 \\
\quad Negligible knowledge & 65 & 20.6 \\
\quad Some knowledge & 164 & 51.9 \\
\quad Moderate knowledge & 55 & 17.4 \\
\quad Extensive knowledge & 6 & 1.9 \\
\bottomrule
\end{tabular}
\end{table}

\subsection{Additional Statistical Results}

\begin{figure}[h]
    \centering
    \begin{subfigure}[t]{0.4\linewidth}
        \centering
        \includegraphics[width=\linewidth]{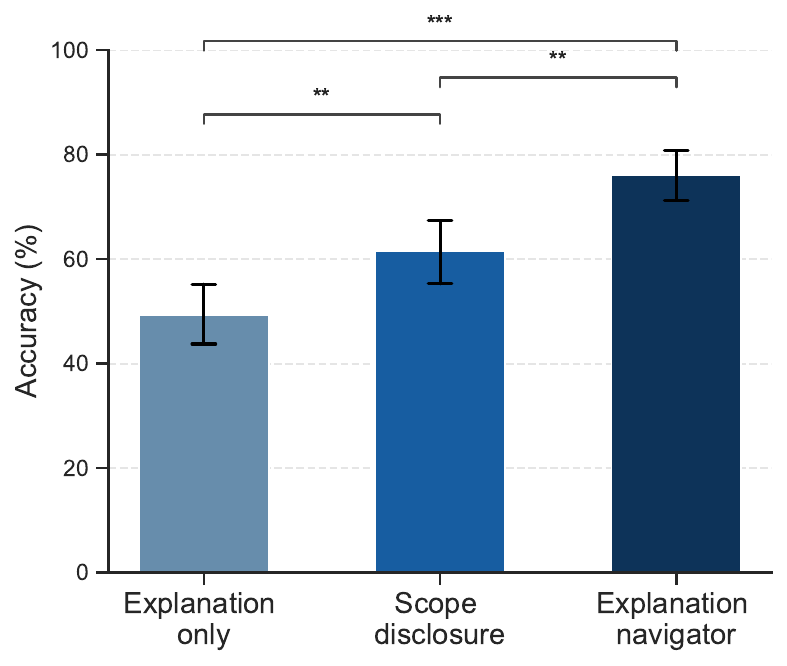}
        \caption{Overall comprehension accuracy for local feature importance.}
        \label{fig:local-overall}
    \end{subfigure}
    \hspace{2em}
    \begin{subfigure}[t]{0.4\linewidth}
        \includegraphics[width=\linewidth]{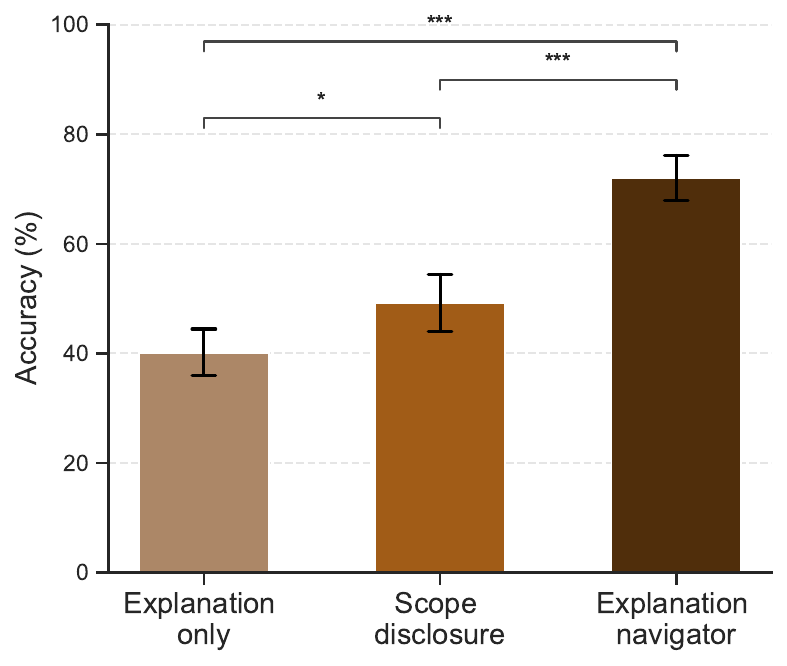}
        \caption{Overall comprehension accuracy for ICE.}
        \label{fig:ice-overall}
    \end{subfigure}
    \caption{Mean overall comprehension accuracy across the interaction conditions for (\subref{fig:local-overall})~local feature importance and (\subref{fig:ice-overall})~ICE. 
    The error bars indicate 95\% confidence intervals. Significance indicators: $^*:p<.05$, $^{**}:p<.01$ and $^{***}:p<.001$.}
    \label{fig:exp-overall-acc}
\end{figure}

\begin{figure}[h]
    \centering
    \includegraphics[width=0.8\linewidth]{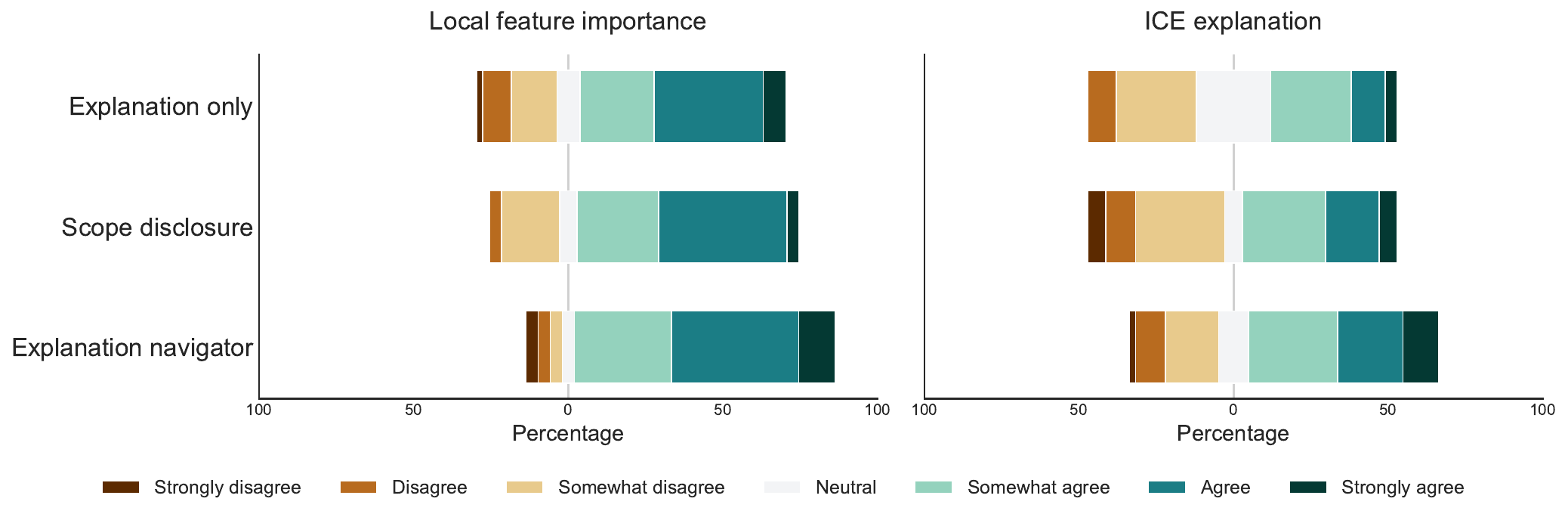}
    \caption{Ratings of the perceived ease of understanding across the interaction conditions for local feature importance (left) and ICE (right) reported by our participants. The ratings were collected on a seven-point Likert scale in response to the following statement: ``This explanation is easy to understand''.}
    \label{fig:sub-easy-understand}
\end{figure}

\end{document}